\documentclass[12pt,a4paper]{article}
\pdfoutput=1
\usepackage{jheppub}
\usepackage[utf8]{inputenc}
\usepackage[T1]{fontenc}
\usepackage{comment}

\usepackage{amsmath,physics,float}  
\usepackage{amssymb}  
\usepackage{latexsym}
\usepackage{graphicx,xcolor}
\usepackage{cleveref}

\usepackage{amsfonts}
\usepackage{braket}
\usepackage{mathrsfs}

\usepackage{subfig} 
\usepackage{empheq}

\usepackage[shortlabels]{enumitem}

\usepackage{pgfplots}
\usepackage{natbib} 
\usepackage{bbm}

\usepackage{bigints}

\newcommand{\beq}{\begin{equation}}
\newcommand{\eeq}{\end{equation}}

\newcommand{\ksi}{\ket{\psi}}
\newcommand{\del}{\partial}
\newcommand{\lc}{\left(}
\newcommand{\rc}{\right)}
\newcommand{\ls}{\left[}
\newcommand{\rs}{\right]} 
\newcommand{\hc}{\text{h.c.}}
\newcommand{\enc}[1]{\left( #1\right)}
\newcommand{\encsq}[1]{\left[ #1\right]}
\newcommand{\T}{\mathcal{T}}

\newcommand{\pdel}{\mathcal{P}_\delta}
\newcommand{\vacket}{\ket{0,\mathcal{S}}}

\newcommand{\pdelta}{\mathcal{P}_\delta}
\newcommand{\pvacex}{\mathcal{P}_0}
\newcommand{\oG}{\text{O}\left(\sqrt{G_N}\right)}
\newcommand{\Cc}{C^\infty_C(\mathbb{R})}
\newcommand{\LtwoR}{L^2\mathbb{R}}
\newcommand{\Cex}{C^{(0)}}
\newcommand{\Cdelta}{C^{(\delta)}}
\usepackage{tikz, pgf}
\usetikzlibrary{decorations.pathmorphing}
\usetikzlibrary{arrows.meta}
\tikzset{%
  >={Latex[width=2mm,length=2mm]},
            base/.style = {rectangle, rounded corners, draw=black,
                           minimum width=4cm, minimum height=1cm,
                           text centered, font=\sffamily},
}
\usetikzlibrary{calc,bending,decorations.markings}
\begin{document}

\title{\boldmath A monogamy paradox in empty flat space}

\author[]{Tuneer Chakraborty,}
\emailAdd{tuneer.chakraborty@icts.res.in}

\author[]{Joydeep Chakravarty,}
\emailAdd{joydeep.chakravarty@icts.res.in}

\author[]{Priyadarshi Paul}
\emailAdd{priyadarshi.paul@icts.res.in}

\affiliation[]{International Centre for Theoretical Sciences (ICTS-TIFR)\\
Tata Institute of Fundamental Research\\
Shivakote, Hesaraghatta,
Bangalore 560089, India.}
\vspace{4pt}

\begin{abstract}
{In a recent paper, Raju showed that the essential features of the monogamy paradox for old flat space black holes could be modeled using a setup in empty AdS, leading to a violation in the monogamy of entanglement there. A physically interesting question is whether such a violation in the monogamy of entanglement can be posed in empty flat space. The answer is not immediately clear since flat space gravity has an entirely different vacuum and infrared structure than the gapped and unique AdS vacuum, which was exploited in Raju's toy model. We answer this question in the affirmative, with an explicit construction. We formulate the paradox in terms of monogamy of CHSH correlations, which we use to quantify the monogamy of entanglement. Extending Raju's analysis to empty flat space, within effective field theory, we show that the entanglement of \textit{approximately local} bulk modes just outside a light cone with modes just inside the light cone as well as with modes situated far away at the past of future null infinity $\lc \mathcal{I}^+_- \rc$ gives rise to an O($1$) violation in the monogamy of entanglement. This cannot be resolved by small corrections of O($\sqrt{G_N}$). The issues arising from the above-mentioned vacuum and infrared features unique to flat spacetime are dealt with by introducing a physically motivated boundary projector onto states below a given infrared cutoff, which allows us to construct suitable operators at $\lc \mathcal{I}^+_- \rc$ that give rise to the violation. We argue that the resolution of the paradox is that our spatially separated observables probe the same underlying degrees of freedom, i.e., such observables act on a non-factorized Hilbert space arising from the Gauss constraint, thereby circumventing the conflict with monogamy of entanglement.}
\end{abstract} 

\maketitle
\pagebreak

\section{Introduction}
Black hole information paradoxes constitute an inter-related web of puzzles that arise due to the existence of the event horizon. These paradoxes have traditionally served as lamp-posts regarding our understanding of various quantum aspects of gravity. Our work is concerned with an important corner of these puzzles: the monogamy paradox in flat space. Originally proposed in \cite{Mathur:2009hf}, the paradox was extensively discussed in \cite{Almheiri:2012rt, Almheiri:2013hfa, Mathur:2012np, Raju:2018zpn, Papadodimas:2012aq, Papadodimas:2013jku, Papadodimas:2015jra, Bousso:2012as, Susskind:2014rva, Mathur:2012jk, Verlinde:2012cy, Nomura:2012sw, Nomura:2012cx, Verlinde:2013uja, Giveon:2012kp, VanRaamsdonk:2013sza, Hutchinson:2013kka, Karlsson:2019vlf, Karlsson:2020uga, Bryan:2016wzx, Nomura:2018kia, Susskind:2012uw, Yoshida:2019qqw, Polchinski:2016hrw, Bousso:2019ykv, Chen:2015gux}. In particular, Raju \cite{Raju:2018zpn} showed that the essential features of the monogamy paradox for old flat space black holes could be modeled using a setup in empty AdS, leading to a violation in the monogamy of entanglement there. The salient point of this setup was that it did not require the existence of a horizon, in contrast to the previous discussions of the paradox. A physically interesting question is whether a violation in the monogamy of entanglement can be described within empty flat space, which resembles our observable universe to a good approximation. Our present work deals with addressing this question.
\subsection{The paradox}
We will briefly discuss the paradox below. Consider an old evaporating black hole in flat space at time $t$, such that $t > t_{\text{Page}}$\footnote{For a black hole with initial entropy given by S which has evaporated to S', the number of states in the \textit{exterior} is approximately $e^{S-S'}$. We thus pose the paradox for old black holes when the exterior contains enough degrees of freedom, i.e., $S' <\frac{S}{2}$. Later on, in line with the principle of holography of information \cite{Laddha:2020kvp, Raju:2020smc, Chowdhury:2020hse}, we will argue why we do not need to necessarily go beyond the Page time in order to set up the paradox, as is done in the standard case.}. The outgoing near-horizon Hawking modes are strongly entangled with the near-horizon interior modes. For the final state to contain all information about the initial state, the near-horizon outgoing modes must also be entangled with Hawking modes that came out of the black hole at early times. However, this situation points to a violation of the monogamy of entanglement, which is an unavoidable consequence of quantum mechanics. This paradox is also closely related to the cloning paradox, which states that within effective field theory, a nice slice can capture both a diary thrown into a black hole and a reconstructed copy of the diary from the outgoing Hawking radiation, thereby violating the no-cloning theorem. The physical picture portrayed by both these paradoxes is that the interior should contain a \textit{copy} of the exterior to resolve contradictions with basic assumptions of quantum mechanics (more precisely, with quantum information theorems).  This picture is reflected within the important idea of black hole complementarity as explored in \cite{THOOFT1985727, Susskind:1993if, Susskind:1993mu, Kiem:1995iy, Banerjee:2016mhh}.

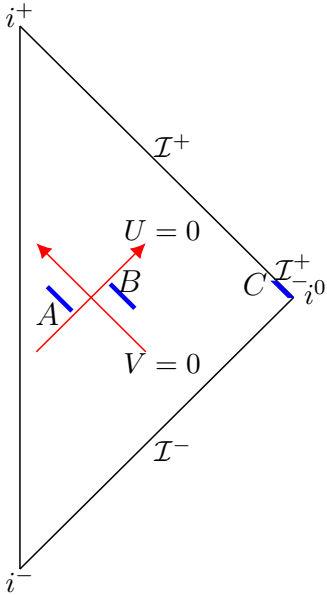
\begin{figure}[H] 
    \centering
    
    \begin{tikzpicture}[scale=0.72] 
    
  \draw [line width=0.2mm] (0,-5) -- (0,5) ;
    \draw  [line width=0.2mm] (5,0)-- (0,-5) ;
    \draw  [line width=0.2mm] (5,0 )-- (0,5) ;

    \draw [->,red,line width=0.2mm,domain=-1:1] plot ({\x+ 1.3}, {1*\x});
    \draw [->,red,line width=0.2mm,domain=1:-1] plot ({\x+ 1.3}, {-1*\x});
    
    \draw [blue,line width=0.6mm,domain=-0.05:-0.5] plot ({\x+ 1}, {-1*\x-0.3});
   \draw [blue,line width=0.6mm,domain=0.05:0.5] plot ({\x+ 1.6}, {-1*\x+0.3});
   \filldraw [blue, fill=blue] (5,0)-- (4.7,0.3 ) -- (4.6,0.3
   )--(4.9,0)-- (5,0);
    
    
    \node (A) at (0.5,-0.3) {$A$};
    \node (B) at (2,0.3) {$B$};
    \node (C) at (4.3,0.25) {$C$};
    
    \node [font=\small] (U0) at (2.6,1.25) {$U = 0$};
    \node [font=\small] (V0) at (2.6,-1.2) {$V = 0$};
    
    \node [font=\small] (chp) at (2.8,2.8) {$\mathcal{I}^+$};
   
     \node [font=\small] (chm) at (2.8,-2.8) {$\mathcal{I}^-$}; 
     
      \node [font=\small] (i0) at (5.4,0.1) {$i^0$}; 
      
      \node [font=\small] (ip) at (0,5.2) {$i^+$};
      
      \node [font=\small] (im) at (0,-5.2) {$i^-$};
      
      \node [font=\small] (chipmm) at (5,0.47) {$\mathcal{I}^+_-$};
\end{tikzpicture}
    \caption{Here the red line $U =0$ denotes a radially outgoing light shell. Regions $A$, $B$, and $C$ are marked in blue. We will study the entanglement of modes smeared over region $A$ with modes smeared over $B$ and $C$ and arrive at a paradox. }
    
    \label{fig:mordor}
\end{figure}

A key idea here is that the monogamy paradox for flat space black holes depends only on the entanglement of near-horizon exterior modes with near-horizon interior modes and also with modes far outside the horizon (e.g., at past of the future null infinity, i.e., $\mathcal{I}^+_-$). Consider the simple situation of a radially outgoing light cone at $r = r_0$ in an empty flat space. A monogamy-type paradox arises here also if we study the entanglement of the modes smeared just inside the light cone (region $A$) with modes smeared just outside the light cone (region $B$) and with another spacelike separated region (region $C$) (See \cref{fig:mordor}). 

\subsection{The toy model in flat space}

We will extend the construction of \cite{Raju:2018zpn} which investigated the monogamy paradox in asymptotically AdS using Bell inequalities to asymptotically flat spacetime to understand the case of old flat space black holes. As done there, we will formulate the paradox using CHSH inequalities \cite{PhysRevLett.23.880}, a convenient restatement of Bell inequalities \cite{PhysicsPhysiqueFizika.1.195}. This formalism allows us to make quantitative statements regarding the monogamy of entanglement \cite{2006quant.ph.11001T, PhysRevLett.87.117901}, in particular, it allows us to rephrase statements about the monogamy of entanglement  in terms of statements regarding the monogamy of CHSH correlations. 

An essential ingredient that facilitates calculations in this setup compared to the original paradox is that the Hamiltonian of gravity is a boundary term \cite{REGGE1974286}, and thus can be used to construct a projector that projects onto the degenerate subspace of vacua labelled by supertranslations. 

\subsubsection{Important features unique to our toy model in flat space}
The vacuum and low energy structure of the Fock space of canonical gravity in flat space is completely different from the same in AdS, due to the presence of supertranslations and the absence of a mass gap. In particular, AdS has a unique vacuum, while the flat space vaccua span a degenerate subspace, and should be specified by their value in the supertranslation sector as well. We build upon previous works \cite{PhysRevLett.46.573, Ashtekar:1981hw, Ashtekar84, Ashtekar:2018lor, Laddha:2020kvp} which have clarified the vacuum structure of flat space, and our definition of relevant operators and their subsequent representation in terms of the Fock space rests on the same. Here, supertranslations are not crucial to setting up the monogamy paradox in flat space but necassarily complicate the rather straightforward analysis in AdS since they introduce an additional vacuum structure. 

Note that in the treatment for the AdS case in \cite{Raju:2018zpn}, there exists a natural cutoff scale set by the cosmological constant. However the issue for flat space gravity in $d=4$ is more complicated, and one needs to specify an infrared cutoff in order to properly define the theory. Therefore in our work, we have introduced a new physically motivated projector $P_{\delta}$, which projects onto energy scales below an infrared cutoff denoted by $\delta$, and utilize the same to construct relevant operators which demonstrate the violation in monogamy of entanglement in our toy model. Physically this means that in practice, we do not work with operators that project onto the vacuum exactly but project onto states with very low energies below an IR cutoff, say $E < \delta$. This also generalizes the more abstract projector onto the vacua introduced in the context of AdS \cite{Banerjee:2016mhh} and in flat space \cite{Laddha:2020kvp}, and their subsequent role in how information is stored at the boundary \cite{Raju:2020smc}.  

While we can setup the monogamy paradox in flat space using the abstract projector introduced in \cite{Laddha:2020kvp} (as we demonstrate), the main thrust of our work is to utilize our physically motivated projector $P_{\delta}$ and use it to set up the monogamy paradox. In our work, given the infrared issues, we firstly demonstrate how operators $C_i$ living in Region C in Fig. \ref{fig:mordor} which have almost the exact correlation with operators $A_i$ in $A$ as operators $B_i$ defined over $B$ have with operators $A_i$ up to O$(\sqrt{G_N})$. Afterwards we then construct them using our physically motivated projector $P_{\delta}$. This requires certain conditions on the smearing functions of relevant observables as we will qualitatively as well as rigorously explain in detail in our work. 

Since these operators in $C$ are constructed such that the $AC$ system has almost identical CHSH correlators as the $AB$ system and consequently the same entanglement, we have a quantum information-theoretic contradiction. 

\subsubsection{Other details about our toy model}

Given that we define our operators $A_i, B_i$ and $C_i$ supported in regions $A, B$ and $C$ respectively, the reader may ask what we mean by operator insertions in a theory of gravity. As opposed to local quantum field theory, there exists no definition of local gauge-invariant operators in a theory of quantum gravity. However, we will work with \textit{approximately local} operators in our case, which involves taking an operator and smearing it over a small spatial interval. One way of thinking about such approximately local operators is to regard them as gauge fixing (i.e. up to small diffeomorphisms) in the bulk. We will not work with explicitly diffeomorphism invariant operators constructed using gravitational dressing, like ones formed by attaching geodesics from the boundary. This is because the paradox necessarily requires us to look at local bulk observables and such constructions are by definition non-local.

We note that the violation in the monogamy is O$(1)$, and cannot be removed by including minor corrections of O$(\sqrt{G_N})$, which are essential to the resolution of Hawking's original paradox and the bags of gold paradox \cite{Maldacena:2001kr, Barbon:2004ce, Anous:2016kss, Fitzpatrick:2016mjq, Chakravarty:2020wdm, Chakravarty:2021tia, Langhoff:2020jqa, Penington:2019kki}. As we discuss below, this is an important observation that strongly hints towards a resolution of the paradox via complementarity, i.e., the interior degrees of freedom are complicated polynomials of the exterior degrees of freedom. The commutators are of O$(1)$ because in principle, we are acting upon the \textit{same degrees of freedom} in the interior and the exterior, and complicated enough exterior operators can probe the information contained in the interior. The resolution is not surprising given that in a theory of gravity, the Hilbert space does not factorize upon spatially partitioning a given manifold. This simple fact follows from the Gauss constraint of gravity. An implicit ingredient that goes into our local quantum field theoretic construction is that the system partitioned into regions $A, B$ and $C$ has factorized Hilbert spaces. However, the monogamy statement here is not violated since upon turning on gravity, the Hilbert space does not factorize, and consequently, the interior and exterior degrees of freedom are not independent degrees of freedom. This is the primary origin of the O$(1)$ violation, which also demonstrates why local quantum field theory is not a good framework to deal with questions regarding quantum information and entanglement in gravity. 

\subsection{Outline of our work}
In \S \ref{chshqft} we review the construction of CHSH operators within quantum mechanics. We generalize this construction in \S \ref{chshlqft} and calculate the CHSH correlation between regions $A$ and $B$ within a local quantum field theoretic framework for asymptotically flat space.

We now list the main non-trivial constructions and results of our work. In \S  \ref{monpar} we outline the construction of the operators $C_i$ living on region $C$, which mimic the action of operators $B_i$ on the global vacuum and write their CHSH correlation with $A_i$'s. We then use the CHSH correlations between $AC$ and $AB$ to set up the paradox in monogamy. In \S \ref{resopar} we argue the resolution of the monogamy paradox in detail. In \S \ref{summarydiscussion} we summarize our work and discuss related perspectives. In Appendix \ref{sec:general_rs_proof}, we give a Fourier analytic proof for the existence of near boundary modes $C_i$ subject to the constraints in our construction.

\section{CHSH inequalities in quantum mechanics}
\label{chshqft}
This section reviews the CHSH inequality for quantum mechanical systems and uses them to provide a factual statement about the monogamy of entanglement.
\subsection{CHSH operator and monogamy of entanglement}
Consider a tripartite system composed of independent subsystems $A, B$ and $C$. We label operators belonging to the algebra of $A$ as $A_i$ and so on for the other subsystems. Let us look at two pairs of operators $A_i$ and $B_i$ where $i \in (1,2)$ which satisfy the commutation relations $\ls A_i, B_j \rs = 0$. These operators are constructed such that their eigenvalues lie in the interval $\ls -1, 1\rs$, or in other words $\norm{A}, \norm{B} \leq 1$. The CHSH operator is given by

\beq
C_{AB} = A_1 B_1 + A_1 B_2 + A_2 B_1 - A_2 B_2.
\eeq

Classically, the maximum value of the CHSH operator is given by 2, which is the case when $A_1$ and $A_2$ are independent while $B_1 = B_2$ or $B_1 = - B_2$. However, this bound no longer holds in quantum mechanics if we evaluate the expectation value of the CHSH operator over a general state $\ksi$. 

In order to estimate the quantum bound on the CHSH operator, let us square the same, which gives us $C_{AB}^2 = 4 - \ls A_1, A_2\rs \ls B_1, B_2 \rs$. Since the norm of the commutator is given by $\abs{\ls A_1, A_2\rs} \leq 2$, we arrive at $\abs{\braket{C_{AB}}} \leq 2\sqrt{2}$.

Now if we consider the square of the expectation value of the CHSH operators defined over $AB$ and $AC$, then the statement of the monogamy of entanglement is as follows \cite{2006quant.ph.11001T}: \beq \label{monoofent}
\braket{C_{AB}}^2 + \braket{C_{AC}}^2 \leq 8.
\eeq
The above relation statement quantifies the maximum entanglement which subsystem $AC$ can possess provided there is a given entanglement among the subsystem $AB$. An interesting conclusion which follows is that there cannot be a scenario where the correlations between $AB$ and $AC$ both possess a non-classical description, i.e. both $\braket{C_{AB}}, \braket{C_{AC}} > 2$. Another outcome is that if the system $AB$ is maximally entangled, i.e. $\braket{C_{AB}} = 2\sqrt{2}$, then $AB$ cannot be entangled. Thus we have a precise statement regarding the violation of monogamy of entanglement, which violates the inequality given in \eqref{monoofent}.

\subsection{Baby example: Bell operators using simple harmonic oscillators}
Consider a pair of commuting simple harmonic oscillators living in separate regions $A$ and $B$. We denote their corresponding annihilation operators as $\alpha_s$, and their respective vacua as $\ket{0}_s$, where $s = A/B$. We want to evaluate the expectation value of the CHSH operator on the thermofield double state where $x^2<1$,
\begin{equation} \label{tfd}
    |\text{TFD} \rangle =\sqrt{1-x^2} \, e^{x \alpha_A^{\dagger} \alpha^{\dagger}_B} \, |0\rangle_A |0 \rangle_B.
\end{equation}
The above state reduces to the standard thermofield double case if we set $x^2 =e^{-\beta}$. Denoting projectors onto the $s$-th vacuum as $P_s$, we now choose Bell operators as follows:
\begin{equation} \label{CHSHops0}
    \begin{aligned}
        \text{$A$ operators:}\quad &
        \begin{aligned}
          & A_1 =  P_A - \alpha_A^{\dagger}P_A \alpha_A 
             \\ & A_2 =   \alpha_A^{\dagger}P_A +P_A  \alpha_A
        \end{aligned}
        \\[12pt]
        \text{$B$ operators:}\quad & B_1= \frac{1}{\sqrt{2}} \big( P_B - \alpha_B^{\dagger}P_B \alpha_B +  \alpha_B^{\dagger}P_B +P_B  \alpha_B \big) \\
        & B_2 = \frac{1}{\sqrt{2}}\big( P_B - \alpha_B^{\dagger}P_B \alpha_B -  \alpha_B^{\dagger}P_B- P_B  \alpha_B \big)
    \end{aligned}
\end{equation}
These operators are inspired by the Bell operators for spin-$\frac{1}{2}$ systems, and might look confusing at first glance. However, expanding the operators in the number basis gives us a much simpler looking form for the same.
\begin{equation}
    \begin{aligned}
        \text{$A$ operators:}\quad &
        \begin{aligned}
          & A_1 =  |0\rangle_A \langle 0 |_A - |1\rangle_A \langle 1 |_A 
             \\ & A_2 =   |0\rangle_A \langle 1 |_A + |1\rangle_A \langle 0 |_A 
        \end{aligned}
        \\[12pt]
        \text{$B$ operators:}\quad & B_1= \frac{1}{\sqrt{2}} \big( |0\rangle_B \langle 0 |_B - |1\rangle_B \langle 1 |_B 
        +  |0\rangle_B \langle 1 |_B + |1\rangle_B \langle 0 |_B \big) \\
        & B_2 = \frac{1}{\sqrt{2}} \big(|0\rangle_B \langle 0 |_B - |1\rangle_B \langle 1 |_B
        -  |0\rangle_B \langle 1 |_B - |1\rangle_B \langle 0 |_B  \big)
    \end{aligned}
\end{equation}
These are precisely the operators used in the spin-$\frac{1}{2}$ problem, with $\ket{0}/\ket{1}$ denoting the two states and the operators resembling combinations of Pauli matrices. We now evaluate the expectation value of the CHSH operator on the thermofield double state, which gives us
\begin{equation} \label{cqm}
\begin{split}
    \langle C_{AB} \rangle & = \sqrt{2}\lc 1+x\rc^3 \lc 1-x \rc.
\end{split}    
\end{equation}
This takes a maximum value at $x = \tfrac{1}{2}$ with the maximum value being $ \langle C_{AB} \rangle = \frac{27\sqrt{2}}{16} \approx 2.39>2$. Therefore using the above construction we see that the thermofield double state is entangled for $x = \frac{1}{2}$, though not maximally entangled.

\section{CHSH inequalities in local quantum field theory}
\label{chshlqft}
In this section, we will extend the above construction of the CHSH correlator for simple harmonic oscillators to analogously construct the CHSH correlator in a local quantum field theory \cite{Raju:2018zpn, Campo:2005sv, summerswerner, PhysRevA.58.4345}. We will then utilize this formalism to calculate $\braket{C_{AB}}$ for smeared modes within a small interval on either side of an outgoing light cone in an empty flat space. This section is computationally intensive, and readers not interested in details of the computation can skip directly to \S \ref{summm}, where we summarize the contents of this section.

\subsection{Basic conventions and choice of operators}

We define Hermitian operators  $(X_s, \Pi_s)$ on the spatially compact regions $A$ and $B$, such that they satisfy canonical commutation relations. Consequently we can also define annihilation operators given by $\alpha_s = \frac{1}{\sqrt{2}}\lc X_s + i \Pi_s \rc$. These operators obey the simple harmonic commutation relations
\beq
\ls \alpha_s, \alpha^{\dagger}_{s'}\rs = \delta_{ss'}.
\eeq
In addition to these modes, there also exist global modes for flat spacetime. These global modes in flat space obey the canonical commutators
\beq \label{cancomm}
[a_{\omega l} , a^{\dagger}_{\omega' l'}] = \delta_{l,l'} \, \delta (\omega- \omega').
\eeq
The global modes are related to $\alpha_s$ by Bogoliubov coefficients
\begin{equation}
    \alpha_s= \sum_{l} \int d\omega\, \lc h_{s}(\omega,l) a_{\omega,l} +  g_{s}^{*}(\omega,l) a_{\omega,l}^{\dagger} \rc,
    \label{bogo}
\end{equation}
where the functions $h_s(\omega,l)$ and $g^{*}_{s}(\omega,l)$ are related by
\beq \label{bogul}
 \sum_{l} \int d\omega\, \ls h_{s}(\omega,l)h^{*}_{s'}(\omega,l) - g_s^{*}(\omega,l)g_{s'}(\omega,l) \rs = \delta_{s,s'}.
\eeq
We rewrite \eqref{bogul} in the following fashion for convenience
\beq
h_{s}. h^{*}_{s'} - g_s^{*}.g_{s'} = \delta_{s,s'},
\eeq
where we have defined $h_{s}. h^{*}_{s'} = \sum_{ l} \int d\omega \, h_{s}(\omega,l)h^{*}_{s'}(\omega,l)$. Let us now consider the scenario where the CHSH correlators are evaluated on the global vacuum state, while the CHSH operators are following combinations of $\alpha_s/\alpha_s^{\dagger}$, which are precisely the same operators in \eqref{CHSHops0}. 
\begin{equation} \label{CHSHops}
    \begin{aligned}
        \text{$A$ operators:}\quad &
        \begin{aligned}
          & A_1 =  P_A - \alpha_A^{\dagger}P_A \alpha_A 
             \\ & A_2 =   \alpha_A^{\dagger}P_A +P_A  \alpha_A
        \end{aligned}
        \\[12pt]
        \text{$B$ operators:}\quad & B_1= \frac{1}{\sqrt{2}} \big( P_B - \alpha_B^{\dagger}P_B \alpha_B +  \alpha_B^{\dagger}P_B +P_B  \alpha_B \big) \\
        & B_2 = \frac{1}{\sqrt{2}}\big( P_B - \alpha_B^{\dagger}P_B \alpha_B -  \alpha_B^{\dagger}P_B- P_B  \alpha_B \big)
    \end{aligned}
\end{equation}
We will proceed using general $\alpha_s$ in \S \ref{vacpro}. Our physical case of interest is described in \S \ref{CHSHAB}, where we will take $\alpha_s$ to be Rindler annihilation modes. Consequently, we have an analogous interpretation of the global state as the thermofield double state as defined in \eqref{tfd}. 
\subsection{Vacuum projector and the most general two-point correlator}
\label{vacpro}

In order to define Bell operators as given in \eqref{CHSHops}, we need to construct projectors onto the ground states of each oscillator, which are given by 
\begin{equation}
     P_s = - \frac{1}{\pi^2}\int_{-\infty}^{\infty} d t_1 \int_{-\infty}^{\infty} dt_2 \int_{0}^{2 \pi} d \theta_s \,  \frac{e^{- (t_1^2+ t_2^2) + \kappa(\theta_s)(t_1 X_s- t_2 \Pi_s)}}{e^{i \theta_s}-1- \epsilon} 
     \label{p0}
\end{equation}
where $\kappa(\theta) \equiv 2 \sqrt{i \tan{\theta}}$ and $\epsilon$ is a small positive constant. The detailed construction of this projector is given in Appendix \ref{p0construction}. 

We can conveniently extract the CHSH correlator from the expression for the \textit{most general two-point correlator}. Using the definition of the projector in \eqref{p0}, the most general two-point function is given by:
\begin{equation}
    \begin{aligned}
      Q[\{v_i,\zeta_i\}] = \tfrac{1}{\pi^4}\int d^2\vec{t} \, d^2 \vec{y} & \int_{0}^{2 \pi} d \theta_A \,  d \theta_B \, \frac{e^{- (\vec{t}^2 + \vec{y}^2)}} {(e^{i \theta_A}-1- \epsilon)(e^{i \theta_B}-1- \epsilon) }\\ & \times \LARGE{\langle}  e^{v_2 {\alpha}^{\dagger}_B} \, e^{ \kappa(\theta_B)(y_1 {X}_B- y_2 {\Pi}_B) } \, e^{\zeta_2 {\alpha}_B} \,
     e^{v_1 {\alpha}^{\dagger}_A} \, e^{ \kappa(\theta_A)(t_1 {X}_A- t_2 {\Pi}_A) } \, e^{\zeta_1 {\alpha}_A} \large{\rangle}
     \label{qfunc}
    \end{aligned}
\end{equation}
where $\vec{t} = (t_1,t_2)$. Let us define $\Tilde{y}_i= \kappa(\theta_B) y_{i}$ and $\Tilde{t}_i= \kappa(\theta_A) t_{i}$ so as to write the expectation value in the above integral as
\beq
        \langle    \mathcal{G} \rangle \equiv \langle e^{v_2 {\alpha}^{\dagger}_B}\, e^{ (\Tilde{y}_1 {X}_B- \Tilde{y}_2 {\Pi}_B) } \, e^{\zeta_2 {\alpha}_B} \,
     e^{v_1 {\alpha}^{\dagger}_A}\, e^{ (\Tilde{t}_1 {X}_A- \Tilde{t}_2 {\Pi}_A) }\, e^{\zeta_1 {\alpha}_A} \rangle
\eeq
The above two-point correlator and its derivatives at $v_i = 0, \zeta_i = 0$ can be used to obtain the correlators of all relevant CHSH operators as defined in \eqref{CHSHops}. As a demonstration, the derivatives of $Q[\{v_i,\zeta_i\}] $ can be easily used to generate correlators of the following form:
\begin{equation}
    \partial_{v_2}^{m_2}\, \partial_{v_1}^{m_1}\, \partial_{\zeta_2}^{n_2} \, \partial_{\zeta_1}^{n_1} \, Q[\{v_i,\zeta_i\}] \big| _{v_{i}= \zeta_i = 0} = \big\langle {\alpha}_B^{\dagger \; m_2} P_B \, {\alpha}_B^{  n_2} \, {\alpha}_A^{\dagger \; m_1} P_A \, {\alpha}_A^{  n_1}  \big\rangle.
\end{equation}
We will write the expression for $\langle    \mathcal{G} \rangle$ in terms of the global modes. This is performed by expressing $\alpha_s$ in terms of global modes using \eqref{bogul}. This computation requires repeated application of the BCH lemma while working in a coherent state basis. The detailed calculation is given in \Cref{app:G}, and we state the final result here.
\begin{equation} \label{mathcalG}
      \langle \mathcal{G} \rangle = \exp\lc \frac{1}{8}\sum\limits_{p,q=1}^{4}  (f _p. f_q^* + f_p^*. f_q)  m_p m_q - \frac{\mathcal{R}}{2} \rc + \text{O} \lc \sqrt{G_N}\rc.
\end{equation}
Here we have added corrections of $\text{O}\lc \sqrt{G_N}\rc$ to include the effects of interactions in an interacting theory of the scalar field coupled to gravity, since the interacting vacuum is different from the global vacuum upto O$\lc \sqrt{G_N}\rc$. Defining $ \zeta^{\pm}_i = \dfrac{(\zeta_i \pm v_i)}{\sqrt{2}}$, the expression for $\mathcal{R}$ is given by
\begin{equation}
    \mathcal{R} =  \big( m_1 \zeta_1^+ + i m_2 \zeta_1 ^- +  m_3 \zeta_2^+ + i m_4 \zeta_2 ^- \big)- v_1 \zeta_1   - v_2 \zeta_2,  
\end{equation}
where the quantities $f_i, m_i $ are defined as:
\begin{subequations}
    \begin{align}
        f_1 & = (h_A + g_A) \; \; ;\; \; f_2 = -i (h_A- g_A) \;\; ;\; \; f_3 = (h_B + g_B) \;\; ;\; \; f_4 = -i (h_B - g_B) \\
          m_1 & = (\Tilde{t}_1+ \zeta_1^+ ) \; \; \;  ; \; m_2 = (- \Tilde{t}_2 + i  \zeta_1^- ) \; \; \; ; \;  m_3 = (\Tilde{y}_1+ \zeta_2^+ ) \; \; \, ;  \;  m_4 = (- \Tilde{y}_2 + i  \zeta_2^- )
    \end{align}
    \label{eq:fm_defs}
\end{subequations}

We can use \eqref{mathcalG} to obtain an expression for $Q[\{v_i,\zeta_i\}] $ in \eqref{qfunc}, since the integrals over $\vec{t}$ and $\vec{h}$ are Gaussian. The $\theta$ integration involves a trivial calculation of the residue in the complex plane. We will not write the expression for general $h_s$ and $g_s$ but will calculate the same for the Rindler to Minkowski Bogoliubov coefficients in the following subsection.

\subsection{CHSH correlation between regions $A$ and $B$ in field theory}
\label{CHSHAB}
Note that our specific case of interest involves smearing operators on bounded regions $A$ and $B$ close to the light cone (See \cref{fig:mordor}). Our smearing choice is such that the operators $\alpha_s$ denote the Rindler oscillators, and $h_s, g_s$ denote the corresponding Rindler to Minkowski Bogoliubov coefficients. While we express the CHSH operators in terms of Rindler operators as given in \eqref{CHSHops}, we take the expectation value in the CHSH correlator over the global vacuum, which is a thermofield double state in terms of the Rindler oscillators. 
\subsubsection{Massless modes in flat space}
Consider a massless scalar coupled to gravity in $d$ dimensional Minkowski space. The modes of the massless scalar end up at future null infinity, a fact that will be important in our posing of the monogamy paradox. The equation for the scalar field is given by:
\beq
\frac{\del}{\del x^{\mu}}\ls \sqrt{-g} \, g^{\mu \nu} \frac{\del \Phi}{\del x^{\nu}} \rs = 0.
\eeq
We solve the above equation in global spherical coordinates (valid for $d\geq 3$): 
\beq \label{metric}
ds^2 = -dt^2 + dr^2+ r^2 d\Omega^2_{d-2}
\eeq
where $\Omega_{d-2}$ denotes the angles of the $d-2$ dimensional sphere. The equation of motion can be solved by putting in the ansatz $\Phi(r,t,\Omega) = T(t) \chi(r) Y_l(\Omega)$, where $Y_l(\Omega)$ denotes spherical harmonics of a $d-2$ dimensional sphere. Here $\chi(r)$ satisfies:
\beq
\frac{d^2 \chi(r)}{dr^2} + \frac{d-2}{r} \frac{d \chi(r)}{dr} + \lc \omega^2 - \frac{l(l+(d-3))}{r^2}\rc \chi(r) = 0,
\eeq
where $\omega $ is the frequency given by:
\beq
\frac{d^2T(t)}{dt^2} = -\omega^2 T(t).
\eeq
The solution for $\chi(r)$ is given by:
\beq
\chi(r) =  \frac{C_1}{r^{\frac{d-3}{2}}} \, J_{m}(r\omega) +  \frac{C_2}{r^{\frac{d-3}{2}}}\, Y_{m}(r\omega)
\eeq
where $J$ and $Y$ denote the standard Bessel functions and $m = l+ \frac{d-3}{2}$. We discard the $Y$ term since it blows up at the origin. Thus the complete solution is given by:
\beq \label{jcb}
\Phi(r,t,\Omega) = K \sum_{ l}  \int d\omega\, a_{\omega, l} \, \frac{J_{m}(r\omega)}{r^{\frac{d-3}{2}}} \, e^{-i\omega t} \, Y_l(\Omega) + \text{h.c.}
\eeq
Here $K$ is a normalization constant used to impose the normalization of the canonical commutator $
\ls a_{\omega, l} , a^{\dagger}_{\omega', l'} \rs = \delta \enc{\omega - \omega'} \, \delta_{l,l'} $. Computing the momenta from the action of the massless scalar and using the equal time canonical commutation relation:
\beq
\ls \Phi(r, t, \Omega), \Pi(r',t,\Omega')\rs = i \delta(r-r') \delta (\Omega_1 - \Omega_2),
\eeq 
we obtain $K = \frac{1}{\sqrt{2}}$. Therefore the scalar field is expressed as
\beq
\Phi(r,t,\Omega) = \frac{1}{\sqrt{2}} \sum_{l} \int d\omega \, a_{\omega, l} \, \frac{J_{m}(r\omega)}{r^{\frac{d-3}{2}}} \, e^{-i\omega t} \, Y_l(\Omega) + \text{h.c.}
\eeq
Note that in the preceding discussion we have suppressed the extra indices of the spherical harmonics. As an example, we can explicitly write them for $d=4$, which gives us
\beq
\Phi(r,t,\theta, \phi) = \frac{1}{\sqrt{2}} \sum_{l,\bar{m}} \int d\omega \, a_{\omega, l} \, \frac{J_{l + \frac{1}{2}}(r\omega)}{r^{\frac{1}{2}}} \, e^{-i\omega t} \, Y^{\bar{m}}_l(\theta, \phi) + \text{h.c.},
\eeq
where we have used $Y^{\bar{m}}_l$ to denote the standard spherical harmonics to avoid confusion with $m$ from \eqref{jcb}. 
\subsubsection{Smeared operators on $A$ and $B$}
We now outline our construction of \textit{approximately local operators} by smearing the scalar field over the bounded interval in such a way that the Rindler modes are extracted out. To perform this, we introduce a tuning function such that it is supported only on the small bounded regions and smoothly dies off. Recall that the regions $A$ and $B$ are situated just inside and outside an outgoing light cone at $r_0$ respectively. Thus we define the smeared operators on the regions $A$ and $B$ by 
\beq \label{smeops}
\begin{split}
    \alpha_A &= \frac{1}{\sqrt{V_\Omega}} \int \frac{dU}{U} \int d^{d-2}\Omega \, r_A^{\frac{(d-2)}{2}} \enc{\frac{U}{U_0}}^{i\omega_0} \T (U) \, \Phi(t_A(U),r_A(U),\Omega)\\
    \alpha_B &= \frac{1}{\sqrt{V_\Omega}} \int \frac{dU}{U} \int d^{d-2}\Omega \, r_B^{\frac{(d-2)}{2}} \enc{\frac{U}{U_0}}^{-i\omega_0} \T(U) \, \Phi(t_B(U),r_B(U),\Omega)\\
       \alpha^{\dagger}_A&= \frac{1}{\sqrt{V_\Omega}} \int \frac{dU}{U} \int d^{d-2}\Omega \, r_A^{\frac{(d-2)}{2}} \enc{\frac{U}{U_0}}^{-i\omega_0} \T^*(U) \, \Phi(t_A(U),r_A(U),\Omega)\\
   \alpha^{\dagger}_B &= \frac{1}{\sqrt{V_\Omega}} \int \frac{dU}{U} \int d^{d-2}\Omega \, r_B^{\frac{(d-2)}{2}} \enc{\frac{U}{U_0}}^{i\omega_0} \T^*(U) \, \Phi(t_B(U),r_B(U),\Omega)
\end{split}
\eeq
Here $r_s$ and $t_s$, $s =A,B$ denote the global spherical coordinates on the regions $A$ and $B$, as given in  \eqref{metric}. The smearing function oscillates increasingly as tend to go near $U=0$, and thus even a small interval near $U=0$ is useful to extract out the Rindler modes. Consequently $U$ is integrated from $U_l$ to $U_h$ such that the tuning function $\T(U)$ vanishes smoothly as it approaches $U_l$ and $U_h$. We work in the  limit $U_0 \to 0$, such that 
\beq
\log \frac{U_l}{U_0} \to -\infty \quad \text{and} \quad \log \frac{U_h}{U_0} \to \infty.
\eeq
 Note that in our convention, we have included the sphere metric determinant $\sqrt{\gamma}$ inside the angular integral in \eqref{smeops}, such that 
$$ V_{\Omega} = \int d^{d-2} \Omega \equiv {\frac{2\pi^{\frac{d-1}{2}}}{\Gamma\lc \frac{d-1}{2}\rc}}.$$
We assume that the errors due to these length scales are of O$(\epsilon)$ such that $\text{O} (\epsilon) \gg \text{O} (\sqrt{G_N})$. In order to impose that the regions $A$ and $B$ remain causally disconnected, we assume the following conditions:
\beq
\begin{split}
t_A (U) = \frac{U}{2} - v_0 \qquad &r_A (U) = r_0 -v_0-\frac{U}{2} \\
t_B (U) = -\frac{U}{2} + v_0 \qquad &r_B (U) = r_0 +v_0+\frac{U}{2}
\end{split}
\eeq
We also impose the following conditions on the tuning function, so that it is sharply centred about a particular frequency $\omega_0$
\beq \label{tuning1}
\T(U) \ls \frac{U}{U_0}\rs^{i\omega_0} = \int \Tilde{\T}(\nu) \ls \frac{U}{U_0}\rs^{i\nu} d\nu,  \qquad \int \frac{d\nu}{\nu} \abs{\Tilde{\T}(\nu)}^2 = \frac{1}{\pi}.
\eeq
using which we can recover the standard expressions for the commutator of the above defined modes $\ls \alpha_s, \alpha_{s'}^{\dagger}\rs = \delta_{ss'}$ (See Appendix \ref{appb} for the detailed calculation). Another relation which will be useful in the computation of $\braket{C_{AB}}$ is
\beq\label{tuva}
\lim_{\nu \to 0} \frac{\tilde{\T}(\nu)}{\nu}=0
\eeq

\subsubsection{Bogoliubov coefficients for Rindler modes}
\label{largenu}
In order to calculate $\braket{C_{AB}}$, we first need to determine the Bogoliubov coefficients so as to calculate the most general two-point correlator, whose simplification has been derived in \eqref{mathcalG}.  Since we have smeared the field over the entire sphere in \eqref{smeops} on either side of the light cone at $r=r_0$, therefore we only need to look at the $l=0$ mode. This is because the modes $l\neq 0$ vanish due to the angular integral. The radial part of the $l =0$ mode takes a very simple form in $d$-dimensions:
\beq
\chi(r) \sim  \frac{J_{\frac{d-3}{2}}(\omega r)}{r^{\frac{d-3}{2}}}
\eeq 
We also note that since we have smeared our operators on very small spatial regions $A$ and $B$, the smearing functions remain almost constant over the region. However using \eqref{tuva} our tuning function vanishes for small frequencies. Consequently the Bogoliubov coefficients in \eqref{bogo} have support only for large frequencies $\omega$, which we denote by $\omega > \omega'$, where $\omega'$ is a large enough frequency above which the Bogoliubov coefficients are non-zero. In the large frequency limit, the above radial function simplifies to
\beq \label{lfl}
\chi (r) \sim \sqrt{\frac{2}{\pi \omega}} \frac{1}{r^{\frac{d-2}{2}}} \cos \lc  \omega r - \frac{(d-2)\pi}{4}\rc
\eeq
Using the large frequency limit, we evaluate the Bogoliubov coefficients. We refer to Appendix \ref{bogorindmink} for the detailed calculation, and state the main result here. 
\beq \label{bogofin}
\begin{split}
    h_A (\omega,0) &= \frac{e^{-i \xi_1}}{2\sqrt{\pi\omega}} \int d\nu\, e^{\pi \nu/2} (\omega U_0)^{-i\nu} \Gamma(i\nu) \tilde{\T}(\nu)  , \\
    g_A^* (\omega,0) &= \frac{e^{i \xi_1}}{2\sqrt{\pi\omega}} \int d\nu\, e^{-\pi \nu/2} (\omega U_0)^{-i\nu} \Gamma(i\nu) \tilde{\T}(\nu)  , \\
    h_B (\omega,0) &= \frac{e^{-i \xi_1}}{2\sqrt{\pi\omega}} \int d\nu\, e^{\pi \nu/2} (\omega U_0)^{i\nu} \Gamma(-i\nu) \tilde{\T}^*(\nu)  ,\\ 
    g_B^* (\omega,0) &= \frac{e^{i \xi_1}}{2\sqrt{\pi\omega}} \int d\nu\, e^{-\pi\nu/2} (\omega U_0)^{i\nu} \Gamma(-i\nu) \tilde{\T}^{*}(\nu).
\end{split}
\eeq

\subsubsection{$\braket{C_{AB}} > 2$ for entangled Rindler modes in flat space}
We will now use the Bogoliubov coefficients given in \eqref{bogofin} to evaluate $\braket{C_{AB}}$, using \eqref{mathcalG}. In order to do this, we need to calculate the $4\times 4$ matrix $f_p \cdot f_q^* + f_p^* \cdot f_q$. The detailed calculation of this matrix is given in Appendix \ref{matrixapp}, and we state the final result.
\begin{equation} \label{ff3}
    f_p \cdot f_q^* + f_q \cdot f_p^* = \frac{2}{1-x^2} \begin{pmatrix}
    1+x^2 & 0 & 2x & 0 \\
    0 & 1+x^2 & 0 & -2x\\
    2x & 0 & 1+x^2 & 0\\
    0 & -2x & 0 & 1+x^2
    \end{pmatrix}.
\end{equation}
Note that the matrix in \eqref{ff3} turns out to be the same as obtained for the Rindler-to-global AdS case in \cite{Raju:2018zpn}. Although solutions to the massless scalar field in AdS and flat space are quite different, it is not surprising that the matrix turns out to be the same. This is because the near-horizon local Rindler modes possess universal features as explained in \cite{Raju:2020smc}.

We now substitute \eqref{ff3} in \eqref{mathcalG} to derive the expression for $\braket{\mathcal{G}}$. In order to do so, we perform the Gaussian integrals over $\vec{t}$ and $\vec{h}$ and evaluate the $\theta$ integral by calculating the residue about the pole. Using the expression for $\braket{\mathcal{G}}$, we derive the result for $\braket{C_{AB}}$, which is again given by
\begin{equation} \label{cqft}
\begin{split}
    \langle C_{AB} \rangle & = \sqrt{2}\lc 1+x\rc^3 \lc 1-x \rc.
\end{split}    
\end{equation}
The reader might ask why our expression for the CHSH operator's expectation value in QFT is precisely the same as the expression we had derived for the quantum mechanical case. This is simply because our chosen operators and states were essentially the same in both cases. Again the expectation value is maximized at $x =\frac{1}{2}$, where $\braket{C_{AB}}$ takes the value
\beq \label{cabfin}
\langle C_{AB} \rangle = \frac{27\sqrt{2}}{16} + \text{O}\lc \sqrt{G_N}\rc + \text{O}\lc \epsilon \rc
\eeq
Here we have included corrections since the interacting vacuum of the scalar-gravity theory is different from the free field vacuum using $\text{O}\lc \sqrt{G_N}\rc$. As defined before, we denote small errors in length scales by $\text{O}\lc\epsilon \rc$.

\subsection{Summary of this section}
\label{summm}
The main goal of this section was to show that within a local quantum field theoretic framework, using a careful choice of operators, we can violate the classical bound. To do this, we first developed the formalism for looking at CHSH correlators in terms of the most general two-point correlator acting on the global vacuum. The key here is to write down the CHSH correlator in terms of Bogoliubov coefficients between the spatially compact regions' modes and the global modes. We then wrote down creation and annihilation operators by smearing the massless scalar field with Rindler smearing functions on small bounded regions $A$ and $B$ situated just inside and outside an outgoing light cone at $r_0$ and calculated the corresponding Bogoliubov coefficients between these operators and the global Minkowski operators. We used these Bogoliubov coefficients to obtain $\braket{C_{AB}}$, where we take the expectation value over the global Minkowski vacuum, which looks like a thermofield double in terms of the Rindler oscillators. In particular, our construction of operators in the local QFT is the same as done for the quantum mechanical case described earlier in \S \ref{chshqft}. Consequently the CHSH correlator is given by \eqref{cqft}, whose maximum value is $\braket{C_{AB}} \approx 2.39$, which violates the classical bound. 

\section{The monogamy paradox in flat space}
\label{monpar}
We will now outline the paradox in the monogamy of entanglement. Previously in \eqref{cabfin} we have derived that up to small corrections, we can obtain $\braket{C_{AB}} = 2.39 > 2$, which indicates non-classicality. We will now consider another region $C$ situated far away from our system $AB$ at $\mathcal{I}^+_-$, and consider operators $C_i$ supported on the same (See \cref{fig:mordor}). Applying \eqref{monoofent} to a local QFT, we have the following upper bound on the CHSH correlators between systems $AB$ and $AC$
\beq \label{monent}
\braket{C_{AB}}^2 + \braket{C_{AC}}^2 \leq 8.
\eeq
In this section, we will show that using the Reeh-Schlieder theorem and the fact that in a theory of gravity, the Hamiltonian is a boundary term \cite{REGGE1974286}, we can create operators $C_i$ such that their action on the vacuum is the same as the action of operators $B_i$. Consequently, the expectation in \eqref{monent} based on local quantum field theory is violated up to an O$(1)$ extent.

Unless indicated otherwise, from here on, we will restrict ourselves to describing the effects of gravity in four dimensions. Firstly we describe the Hilbert space of the theory and construct operators relevant to our calculation. Then we calculate the $\braket{C_{AC}}$ correlator and pose the paradox. We will further discuss conditions on the vacuum structure under which we can similarly pose the paradox in general dimensions. Towards the end of this section, we discuss the resolution of the paradox.

\subsection{Gravity in asymptotically flat spacetime}
 
In this subsection, we will describe the Hilbert space of the four-dimensional flat space theory and construct a boundary projector onto low energy states. This projector will be essential to construct bounded operators $C_i$ within a small region at the past of future null infinity ($\mathcal{I}^+_-$). Readers familiar with the details of this section can directly proceed onto \S \ref{caccac}.
\subsubsection{The Hilbert space}
 A good coordinate system which encapsulates the asymptotic large-$r$ structure near the future null infinity is the retarded Bondi coordinates \cite{Bondi:1960jsa}. 
\beq
ds^2 = -du^2 - 2\,du dr + r^2 \gamma_{AB}\, d\Omega^A d\Omega^B + r \, C_{AB}\, d \Omega^A d\Omega^B + \frac{2 m_B}{r} du^2 + \gamma^{DA}D_D C_{AB} \, du \, d\Omega^B + \dots
\eeq
There is an infinite-dimensional symmetry group in the asymptotic region consistent with the leading falloff given above \cite{Bondi:1960jsa, Bondi:1962px, Sachs:1962zza, Sachs:1962wk, Strominger:2017zoo, Compere:2018aar}. These symmetries are called supertranslations which are generated by the following charges:
\beq
Q_{lm} = \frac{1}{4\pi G_N} \int \sqrt{\gamma} \, d^2\Omega \, m_B (u = -\infty, \Omega) \, Y_{l,m}(\Omega)
\eeq
 The Bondi news is given by the $u$-derivative of the shear, $N_{AB} = \partial_u C_{AB}$. This tensor has a zero mode, which is used to split the supertranslation charges into two parts, a soft part and a hard part. Technically it is the soft part that leads to the asymptotic symmetries, while the hard part contains stress-energy contributions.

 We will briefly talk about the Fock space of this asymptotic theory \cite{PhysRevLett.46.573, Ashtekar:1981hw, Ashtekar84} which is elaborated in more detail in \cite{Ashtekar:2018lor, Laddha:2020kvp}. Since the news tensor contains a zero mode, the vacuum must be specified not only by the annihilation of the positive frequency modes of the news tensor and the scalar field but should also be labelled by the eigenvalue under the supertranslation sector
\beq \label{fsp1}
Q_{lm}\ket{0, \{s\}} = s_{l,m} \ket{0, \{s\}}
\eeq
Here $0$ denotes that the positive frequency modes of the field, i.e. the hard part of the supertranslation charges annihilate the vacuum. By smearing over the energies using suitable tuning functions\footnote{To see why smearing over energy kets is convenient, note that in QFT without the inclusion of gravity, the vacuum is normalizable but the excited states are not. In order to work with states satisfying nice properties, we smear over energy kets there as well. Also note that the Fock space in both cases, i.e., QFT with or without gravity is separable.}, where the smearing scale can be taken to be arbitrarily small, the inner product between two states is given by
\beq \label{fsp2}
\braket{\, \{ n_{\omega} \}, \{ s\}\, |\, \{ n'_{\omega} \}, \{ s'\} \,} =\prod_{l,m} \delta_{\{ n_{\omega} \},\{ n'_{\omega} \}} \, \delta \lc s_{l,m} - s'_{l,m}\rc.
\eeq
where the Dirac delta function goes over the space of all $l,\, m$. Consequently, we can build up the Hilbert space by acting with the massless scalar and the news field on the top of each vacuum labelled by $\ket{0, \{s \}}$. Thus the Hilbert space is fragmented into different sectors, with an element from one sector orthogonal to another from a different sector. Thus Hilbert space of canonical gravity is given by
\beq \label{fsp3}
\mathcal{H} = \bigoplus_{\{ s \}} \mathcal{H}_{\{ s \}} 
\eeq
We will pause here to clarify some important aspects while working with the Fock space as described in \eqref{fsp1}, \eqref{fsp2} and \eqref{fsp3}. Physically in order to compute meaningful quantities, we write the state of our scalar field as follows:
\beq \label{smket}
\ket{\{ n_{\omega} \}, \mathcal{S}} \equiv \int \lc \prod_{l,m} ds_{l,m}\rc \mathcal{S}(\{ s\}) \ket{\{ n_{\omega} \}, \{ s\}},
\eeq
where we have smeared the supertranslation of the vacuum, with the peak of the smearing function $\mathcal{S}(\{ s\})$ centred about a particular $s_{l,m}$ to ensure normalizability of states. The smearing function $\mathcal{S}(\{ s\})$ is chosen such that our states have unit norm. Therefore using \eqref{smket} and \eqref{fsp2}, the inner product between smeared states is given by  
\beq \label{smkett}
\braket{\, \{ n_{\omega} \}, \mathcal{S}\, |\, \{ n'_{\omega} \}, \mathcal{S}' \,} = \delta_{\mathcal{S}, \mathcal{S}'} \delta_{\{ n_{\omega} \},\{ n'_{\omega} \}}.
\eeq
 
We note a critical assumption in our discussion: we have ignored UV corrections, e.g., stringy effects, and assumed that the low energy effective physics correctly describes the low energy structure of quantum gravity. This assumption seems quite reasonable since gravity is an excellent effective field theory up to the Planck scale. In our work, we pose the paradox within a low energy framework where we perform only tree-level calculations, and hence we are not bothered by any possible modification to the Hilbert space introduced by a UV completion of gravity such as string theory. 

Finally, we also note that our construction manifestly ensures that our Fock space is separable. This statement can also be motivated using constructive QFT \cite{Streater:1989vi, Haag:1992hx, BaezSegalZhou+2014, doi:10.1142/97818481602240007}.

\subsubsection{Boundary projector}
We will now use the gravity Hamiltonian to write down a projector in asymptotically flat spacetime \cite{Laddha:2020kvp}. We first write the Bondi mass, which is the integration of the Bondi mass aspect over the sphere at infinity. 
\beq
M(u) = \int d^2\Omega \,\sqrt{\gamma} \, m_B(u, \Omega) 
\eeq
 Note that the Bondi mass at $u\to -\infty$ is the $m=0,\, l=0$ component of supertranslation charges $Q_{lm}$. The Bondi mass reduces to the canonical ADM Hamiltonian in the limit $u\to -\infty$ \cite{PhysRevLett.43.181, Arnowitt:1962hi, REGGE1974286}:
\beq
\lim_{u\to -\infty} \frac{M(u)}{4\pi G_N} = H.
\eeq
The ADM Hamiltonian can be expressed in terms of the boundary metric, which is given by
\beq
 H = \lim_{u\to -\infty} \frac{M(u)}{4\pi G_N} = \lim_{r\to \infty} \frac{1}{4\pi G_N} \int d^2\Omega \, \sqrt{\gamma} \lc r \, h_{00}(r, \Omega) \rc. 
\eeq
Using this boundary Hamiltonian, we can write down a projector residing at $\mathcal{I}^+_-$. The projector onto the subspace of vacuum states labelled by supertranslations is constructed by taking the following limit \cite{Laddha:2020kvp,Banerjee:2016mhh}:
\beq \label{abspro}
\mathcal{P}_{0} = \lim_{a \to \infty} \exp\lc- a H \rc.
\eeq
where the subscript $0$ in the projector denotes that we are projecting onto the degenerate subspace of zero energy states spanned by supertranslations. We can express this projector as an operator on the Fock space as follows\footnote{Throughout this work, we will use the basis $\lc \{ s\} \rc $ to denote the supertranslation elements within projectors rather than the basis of smeared supertranslations $\lc \mathcal{S}\rc$ to do so. The smeared basis is utilized while labelling the vacuum state denoting our system. }
\beq \label{pzero}
\mathcal{P}_{0} = \int \lc \prod_{l,m} ds_{l,m} \rc  \,  \ket{0, \{ s\}} \bra{0, \{s \}} + \text{O}\lc \sqrt{G_N}\rc.
\eeq
However, in practice, the projector written in \eqref{abspro} is defined only in an abstract sense. A more physically motivated projector in flat space should project only up to energies below an IR scale $\delta$, such that O$(\delta) \gg \text{O} (G_N)$. We should be able to set the IR cutoff $\delta$ arbitrarily small, i.e.; it should not appear in answers to a well-defined physical problem. The expression for the projector onto low energy states in the Fock space is given by
\beq
\pdel = \Theta \lc \delta - H \rc.
\eeq
Since our projector is a function of the Hamiltonian, it is given by a boundary term as well. The representation of this operator over states labelled by the energy and supertranslations is given by
\beq \label{pdelta}
\pdel = \int \lc \prod_{l,m} ds_{l,m} \rc \sum_i \,\Theta \lc \delta - E_i \rc \, \ket{E_i, \{ s\}} \bra{E_i, \{s \}} + \text{O}\lc \sqrt{G_N}\rc,
\eeq
where for notational convenience we have relabelled the states as $\ket{E_i, \{ s\}}$. However it should be kept in mind that states satisfy the inner product in \eqref{smkett}. In particular states with different energy distributions but with same total energy should be thought of as labelled by different values of the index $i$.

\subsection{CHSH correlation between regions $A$ and $C$}
\label{caccac}
 We note that our calculation of $\braket{C_{AB}}$ in the absence of gravity remains unmodified when we turn on gravity (up to O($\sqrt{G_N}$)) since we have simply fixed $s_{l,m}$ in \eqref{smket}. Physically, our operator insertions within $C_{AB}$ are hard, and such operator insertions do not change the soft quantum numbers. As a result, the calculation of $C_{AB}$ goes through in gravity.
 
 Now we can construct a spacelike nice slice containing the regions $A, B$ and $C$. On this slice, using the Reeh-Schlieder theorem \cite{Reeh:1961ujh, Witten:2018zxz} we can construct local operators $Q_i$ living on the region $C$ which replicate the action of hard operators living on region $B$, such that
\beq \label{qeye}
Q_i \ket{0, \{s\}} = B_i \ket{0, \{ s\}} + \text{O} \lc \sqrt{G_N} \rc
\eeq
where as usual we have added contributions due to the interacting vacuum. Apart from the theorem guaranteeing their existence, in $d$-dimensions the operators $Q_i$ can be explicitly constructed as follows. 

Consider region $C$ denoted by the Rindler wedge covered by the chart $z= Z+ \zeta\cosh\tau$, $t=\zeta\sinh\tau$, in the domain $0 < \zeta <\infty$, $-\infty<\tau<\infty$ so as to have $z>Z+\abs{t}$. To make this wedge spacelike separated from the region $AB$ we keep $Z\gg r_0$. The metric is 
\begin{equation}
    ds^2= -\zeta^2d\tau^2 + d\zeta^2 + \sum_{i=1}^{d-2} dx_i^2.
\end{equation}
We take a separable solution of the form $\Phi(\tau,\zeta,\mathbf{x})=e^{-i\enc{\omega\tau-\mathbf{k}\cdot \mathbf{x}}} \chi(\zeta)$ in order to solve $\Box \Phi=0$. The $\zeta$-equation is given by
\begin{equation}
    \zeta^2\frac{d^2\chi}{d\zeta^2} + \zeta \frac{d\chi}{d\zeta} +  \enc{\omega^2-k^2\zeta^2}\chi=0,
\end{equation}
where $k\equiv\abs{\mathbf{k}}$. Imposing boundedness of the solution in the limit $\zeta\to\infty$ at fixed $\tau$ and $\mathbf{x}$, the the field can be expressed as
\begin{equation}
    \Phi(\tau,\zeta,\mathbf{x}) = \int_{\omega>0} \frac{d\omega d\mathbf{k}}{\enc{2\pi}^{\frac{d-1}{2}}}\, \sqrt{\frac{2}{\omega}}  b_{\omega,\mathbf{k}} e^{-i\enc{\omega\tau-\mathbf{k}\cdot\mathbf{x}}} \frac{K_{i\omega}\enc{k\zeta}}{\abs{\Gamma(i\omega)}} +\hc.
\end{equation}
where $K_{i\omega}(k\zeta)$ is the modified Bessel function of the second kind. The $\omega$ dependent factors inside the integral ensure the canonical commutation relations\footnote{We make use of \begin{align*}
    \int_{0}^\infty \frac{dx}{x}\, K_{i\omega}(x) K_{i\omega'}(x) &= \frac{\pi}{2}\abs{\Gamma (i\omega)}^2 \delta(\omega-\omega') \\ 
    \int_{0}^\infty  \frac{d\omega}{\abs{\Gamma(i\omega)}^2} \,K_{i\omega}(x) K_{i\omega}(x') &=\frac{\pi}{2}x\delta(x-x')
\end{align*}}: $[b_{\omega,\mathbf{k}},b^\dagger_{\omega',\mathbf{k}'}]=\delta(\omega-\omega')\delta(\mathbf{k}-\mathbf{k}')$ and $[b_{\omega,\mathbf{k}},b_{\omega',\mathbf{k}'}]=0$ \cite{Papadodimas:2012aq}. 

On the complement of this Rindler wedge we can again write down Rindler-like coordinates, where in addition to the crossed over modes $b$ and $b^{\dagger}$, there also exist a set of modes with support on $z<Z$ at $t =0$ denoted by $\tilde{b}$ and $\tilde{b}^{\dagger}$. Since these tilde operators are spacelike to Rindler wedge operators, they commute. Thus within the complement of the Rindler wedge, where the coordinates are $t=-\zeta \sinh \tau, z=Z-\zeta\cosh\tau$, the field operator can be written as
\begin{equation}
    \Phi(\tau,\zeta,\mathbf{x}) = \int_{\omega>0} \frac{d\omega d\mathbf{k}}{\enc{2\pi}^{\frac{d-1}{2}}}\, \sqrt{\frac{2}{\omega}}  \tilde{b}_{\omega,\mathbf{k}} e^{i\enc{\omega\tau-\mathbf{k}\cdot\mathbf{x}}} \frac{K_{i\omega}\enc{k\zeta}}{\abs{\Gamma(i\omega)}} +\hc
\end{equation}
This is precisely how the smeared operators $A_i$ and $B_i$ in the previous section can be constructed from wedge operators and its complement. From the Bisognano-Wichmann construction \cite{Witten:2018zxz}, the complement operators are related to the wedge operators as:
\beq
\tilde{b}_{\omega, l}\ket{0, \{s\}} = e^{-\pi \omega} b^{\dagger}_{\omega, l}\ket{0, \{s\}}; \qquad \tilde{b}^{\dagger}_{\omega, l}\ket{0, \{s\}} = e^{\pi \omega} b_{\omega, l}\ket{0, \{s\}}
\eeq
where $\ket{0, \{s\}}$ denotes the global vacuum. Thus we systematically obtain \eqref{qeye}. Using this construction, the action of the complement operators $B_i$ on the vacuum can be written in terms of the action of the wedge operators $Q_i$ on the vacuum.

The operators $Q_i$ constructed above are in general unbounded, whereas in order to calculate CHSH correlations we require bounded operators. We will now construct operators $C_i$ such that they satisfy
\beq \label{boundc}
\norm{C_i} = \braket{B_i^2} + \text{O} \lc \sqrt{G_N} \rc \qquad \braket{A_j C_i} = \braket{A_j B_i} + \text{O} \lc \sqrt{G_N} \rc
\eeq
firstly using the projector $\mathcal{P}_{0}$ onto the flat vacua subspace. We will then use the physical projector $\pdel$, and show that there exist our required operators $C_i$, and construct them explicitly. 

\subsubsection{Construction of $C_i$ using $\mathcal{P}_{0}$}
\label{cpzero}
In this part, we outline the construction of operators $C_i$ using the exact projector onto the vacuum. For notational simplicity, we will suppress factors of $\text{O} \lc \sqrt{G_N} \rc$ within this subsection, and will reinstate the same in \S \ref{messi}.

In order to construct bounded operators from $Q_i$, we take combinations of products of $Q_i$ with the projector $\mathcal{P}_{0}$. Consequently we recover the action of $B_i$ on the vacuum, and therefore the resulting operator can be bounded. We define the operators $C_i$ by the following expression
\begin{equation} \label{exop}
    C_i \equiv \frac{\braket{B_i^2}\enc{Q_i\mathcal{P}_{0}+ \mathcal{P}_{0} Q_i^\dagger- \braket{B_i}\mathcal{P}_{0}} - \braket{B_i}Q_i \mathcal{P}_{0} Q_i^\dagger}{\braket{B_i^2}-\braket{B_i}^2},
\end{equation}
where the cumulants are defined with respect to the smeared state $\ket{0, \mathcal{S}}$. The operators constructed in \eqref{exop} might appear out of the blue, however they are systematically constructed by considering the subspace spanned by  $\{ \ket{0, \{s\}}, \,  B_i \ket{0, \{s\}}\}$. For notational convenience, we also define 
$$\ket{B_i, \{s\}} \equiv B_i\ket{0, \{s\}} \quad \text{and} \quad \beta_i \equiv \sqrt{\braket{B_i^2} - \braket{B_i}^2}.$$
Then the construction of $C_i$ is as follows. We start with a candidate $C_i$ with linear combination of all possible outer products which do not involve cross terms from different superselection sectors, i.e.:
$$\{\ket{ 0, \{s\}} \bra{0, \{s\}}; \, \ket{0, \{s\}} \bra{B_i, \{s\}}; \ket{B_i, \{s\}} \bra{0, \{s\}}; \,\&\, \ket{B_i, \{s\}} \bra{B_i, \{s\}}\}$$
multiplied by undetermined coefficients. These coefficients can be systematically determined such that they satisfy the bounds in \eqref{boundc}, which gives us \eqref{exop}. To demonstrate this, we rewrite the expression for $C_i$ in \eqref{exop} as a linearized sum of outer products with determined coefficients
\begin{equation} \label{starwars}
\begin{split}
    C_i =& \int \lc \prod_{l,m} ds_{l,m} \rc \frac{\braket{B_i^2}}{\beta_i^2} \lc\ket{B_i,\{s\}}\bra{0,\{s\}} \ls 1- \int \lc \prod_{l,m} ds'_{l,m} \rc \frac{\ket{B_i,\{s'\}}\bra{B_i,\{s'\}}}{\braket{B_i^2}}\rs \rc \\
    &+ \int \lc \prod_{l,m} ds_{l,m} \rc \frac{\braket{B_i^2}}{\beta_i^2} \lc \ket{0,\{s\}}\bra{B_i,\{s\}} \ls 1- \int \lc \prod_{l,m} ds'_{l,m} \rc \ket{0,\{s'\}}\bra{0,\{s'\}}\rs\rc.
\end{split}    
\end{equation}
 The proof of boundedness of $C_i$ as defined in \eqref{exop} and $\braket{A_j C_i} = \braket{A_j B_i} + \text{O} \lc \sqrt{G_N} \rc$ is given in Appendix \ref{appc}. 

\subsubsection{Construction of $C_i$ using $\pdel$}
\label{422}
We will now proceed with the construction of $C_i$ using the more physical projector $\pdel$. Motivated by \eqref{exop}, we can write a similar expression for $C_i$, which is valid up to an O$(\epsilon)$ correction. 
\begin{equation} \label{exop2}
    C_i \equiv \frac{\braket{B_i^2}\enc{Q_i\pdel+ \pdel Q_i^\dagger- \braket{B_i}\pdel} - \braket{B_i}Q_i \pdel Q_i^\dagger}{\braket{B_i^2}-\braket{B_i}^2}.
\end{equation}
To see why operators in \eqref{exop2} are valid operators upto O$(\epsilon)$, we firstly decompose the projector $\pdel$ as
\beq \label{deltapmath}
\pdel = \mathcal{P}_{0} + \delta \mathcal{P}.
\eeq
The claim holds provided the contribution to $C_i$ arises solely due to $\mathcal{P}_{0}$, with $\delta \mathcal{P}$ not contributing to $C_i$. By acting operators $C_i$ on the vacuum $\ket{0, \mathcal{S}}$, we can ensure that the chief contribution to $C_i$ comes from $\mathcal{P}_{0}$ by demanding 
\beq \label{refeeq}
\abs{\braket{0, \mathcal{S}|\, Q_i\,|E_j, \mathcal{S}}}  \sim \text{O} \lc \epsilon \rc \quad  \text{and} \quad 
\abs{\braket{0, \mathcal{S}|\, A_i \, Q_k \,|E_j, \mathcal{S}}}  \sim \text{O} \lc \epsilon \rc .
\eeq
where the net energy $E_j$ of the state satisfies $0<E_j < \delta$. This renders $\delta \mathcal{P}'$s contribution within $C_i$ very small, and consequently the $C_i$'s defined in \eqref{exop2} satisfy the constraints in \eqref{boundc}. Note that these extra contributions arise due to the last term in \eqref{exop2}.

Note here that O$\lc \epsilon \rc$ denotes minor errors introduced due to smearing scales, i.e., the operator smearing and the wedge smearing scales. We group all such scales as O$\lc \epsilon \rc$ since relatively these errors are of the same magnitude, in contrast to much more minor errors of O$\lc \sqrt{G_N}\rc$.

Under what condition can we ensure \eqref{refeeq}? To begin, consider a single-particle state $\ket{j_{\Omega}} = a_{E_j}^{\dagger} \ket{0, \mathcal{S}}$, such that $E_j < \delta$. To ensure \eqref{refeeq}, we first evaluate the expression $Q_1 \ket{j_{\Omega}}$.
\beq
\begin{split} \label{commuter}
Q_1 \ket{j_{\Omega}} &= \ls Q_1, a^{\dagger}_{E_j}\rs \ket{0, \mathcal{S}} +  a^{\dagger}_{E_j} Q_1 \ket{0, \mathcal{S}}
 \end{split} 
\eeq
We will now argue that both the terms in \eqref{commuter} can be set small enough, thereby satisfying the conditions in \eqref{refeeq}. To see why the first term is small, let us discuss the energy scales in the problem. Apart from the Planck scale, there are two other energy scales in the problem: the energy $\omega'$ as defined in \S \ref{largenu} (below which the Bogoliubov coefficients were close to zero); and $\delta$, which denotes the IR cutoff. Now recall that $B_1$ is given by
\beq \label{fddd}
B_1= \frac{1}{\sqrt{2}} \big( P_B - \alpha_B^{\dagger}\,P_B\, \alpha_B +  \alpha_B^{\dagger}\,P_B +P_B\, \alpha_B \big)
\eeq
where $P_B$ denotes the projector onto the $B$-vacuum, i.e., $P_B = \ket{0_B}\bra{0_B}$, and where we have suppressed the supertranslation labels for convenience. Note that the modes $\alpha_B$ in \eqref{fddd} are related to the global modes as given in \eqref{bogo}, and consequently the vacuum $\ket{0_B}$ is related to the global vacuum $\ket{0}$ as follows:
\beq \label{fdddd}
\ket{0_B} = \exp \lc \sum_{jk} \frac{1}{2}\, a^\dagger_j\,  C_{jk} \, a^\dagger_k\rc  \ket{0}.
\eeq
where $C_{jk}$ is the matrix outlined in the footnote\footnote{In general, for modes related by Bogoliubov transformations 
\beq
a_i = \sum_{i,j} \alpha_{ij}\, b_j + \beta_{ij}\,b^{\dagger}_j
\eeq 
with $a_i\ket{\Omega} = 0$ and $b_j\ket{X} = 0$, the vacua $\ket\Omega$ and $\ket{X}$ are related by:
\beq
\ket{\Omega }= \exp \lc \frac{1}{2}\, b^\dagger_j\,  C_{jk} \, b^\dagger_k\rc  \ket{X}
\eeq
where the matrix $C_{jk}$ is given by
\beq
C_{mj} = - \sum_i \beta^*_{mi}\, \gamma_{ij} \quad \text{with} \quad \sum_i \alpha_{ji} \, \gamma_{ik} = \delta_{ik}
\eeq
where $\delta_{ik}$ denotes the Kronecker delta function.
}. Thus the operators $B_i$ can be expressed in terms of the global modes as outlined above. Note that the global operators $a_i^{\dagger}$ can be constructed only if we have access to the entire spacelike slice $\Sigma$, i.e.:
\beq
a_k^{\dagger} = \int_{\Sigma} \phi(x)\,  e^{+ikx}\,  \frac{d^{d-1}x}{(2\pi)^{d-1}}
\eeq
and consequently $B_i$ can only be written down provided we have access to the whole entire spacelike slice. However, since we have access only to the wedge and not the entire slice, an exact wedge reconstruction of the operator $B_i$ is impossible. In particular, any attempt to reconstruct $a_k^{\dagger}$ will also necessarily include other creation and annihilation operators.
\beq \label{keita}
\int_{\Sigma'} \phi(x)\,  f_k(x)\,  \frac{d^{d-1}x}{(2\pi)^{d-1}} = a_k^{\dagger} + \sum_j c_j \, a_j + \sum_{j\neq k} d_j \, a_j^{\dagger}
\eeq
 where $\Sigma' \in \Sigma$ denotes the spacelike part of the wedge and where $f_k(x)$ is a smearing function with support on $\Sigma'$. In spite of this obstruction, the Reeh Schlieder theorem, and in particular our wedge reconstruction analysis in \S \ref{caccac} gives us \eqref{qeye}, i.e.: 
\begin{equation*}
  Q_1 \ket{0, \{s\}} = B_1 \ket{0, \{ s\}}  
\end{equation*}
The critical point here is that there exist smearing functions $f_k(x)$, with support on the wedge, which convolves with field operator $\phi(x)$ using which we can construct such an operator $Q_1$ from the wedge. Then the practical way to construct $Q_1$ is as follows: we attempt to \textit{closely simulate} $B_i$ by wedge reconstructing the global creation and annihilation operators as in \eqref{keita}. We perform this attempt by choosing wedge smearing functions appropriately and substituting the closely simulated operators in \eqref{fddd} (which is essentially an infinite string of creation and annihilation operators from \eqref{bogo} and \eqref{fdddd}). Consequently, we have a vast choice in choosing the smearing functions since each global operator insertion in \eqref{fdddd} can be simulated using a reconstructed wedge operator. This method gives us the action of $B_1$ on the vacuum using $Q_1$. As a result, $Q_1$ has additional terms than $B_1$ since we cannot precisely reconstruct the operator $B_1$.

 Upon normally ordering, $Q_1$ takes the following form:
\beq \label{fdd}
Q_1 = B_1 + \sum_j p_{1j}( a_k^{\dagger}) + \sum_j q_{1j} (a_k)
\eeq
where $q_{1j}$ contains at least one annihilation operator, while $p_{1j}$ contains the remaining terms with zero or more creation oscillators (Note that the operator $\ket{0}\bra{0}$ inside \eqref{fddd} cancels the remaining terms).
In order to demand \eqref{qeye}, the complex coefficients multiplying operator distributions inside $p_{1j}$ in \eqref{fdd} are conveniently adjusted using smearing functions such that the following inner product is ensured:
\beq \label{fd6}
\braket{0,\mathcal{S}|\,Q_1^\dagger \, Q_1|0,\mathcal{S}} \approx \braket{0,\mathcal{S}|\,B_1^\dagger \, B_1|0,\mathcal{S}}.
\eeq
Now using \eqref{bogo}, \eqref{fddd} and \eqref{fdddd}, we will argue that the wedge reconstructed operators $Q_1$ in \eqref{fdd} satisfy \eqref{qeye} along with ensuring the first term in \eqref{commuter} is small enough, provided that $B_i$ and the smearing functions $f_k(x)$ satisfy the following conditions:

\begin{enumerate}
    \item $B_1$ has a small overlap with $a^{\dagger}_{E_j}$. In other words, the modes constituting $B_1$ are sufficiently high energy modes which are constructed such that $\omega' \gg \delta$. This ensures that 
    \beq \label{fd0}
    \ls B_1, a^{\dagger}_{E_j} \rs \sim \text{O}(\epsilon)
    \eeq
     Here O$(\epsilon)$ denotes the order of overlap between the high energy and the low energy modes due to smearing scales. A physically intuitive way to understand why we require $\omega' \gg \delta$ in flat space is to view the projection onto states below $\delta$ as noise in our description over the ground state subspace. Naturally, we do not want operator insertions inside the correlators characterized by frequencies within the noisy regime, rendering measurements meaningless. Therefore the noise $\delta$ needs to be set sufficiently low enough for the construction to work\footnote{Note that in $AdS$, the $AdS$ radius sets a natural length scale restricting $\delta$ once and for all, which is not the case here since the cosmological constant is zero.}.
     
      We also note that if $\omega' < \delta$, the first term in \eqref{fd0}  cannot be $\text{O} (\epsilon)$, but constitutes an $\text{O} (1)$ contribution. 

\item
This leaves us with the third term (note that the second term commutes with $a^{\dagger}_{E_l}$, and also has a very small magnitude) i.e., an infinite number of annihilator strings $q_{1j}$. These can provide a large contribution to the commutator in \eqref{commuter}. To circumvent this, we require that our smearing functions is chosen such that the following contribution is ensured:
\beq \label{fddddd}
\sum_j \ls  q_{1j} (a_k), \,  a^{\dagger}_{E_l} \rs  \sim \text{O}(\epsilon).
\eeq
In particular, the constraint in \eqref{fddddd} implies that the third term of the operator $Q_1$ given in \eqref{fdd} has a minimal contribution from annihilators below $\delta$, and hence a slight overlap.
\end{enumerate}
Thus using the conditions \eqref{fd6}, \eqref{fd0} and \eqref{fddddd} on $Q_1$, we can ensure that the modes constituting $Q_1$ are engineered such that the following commutator in \eqref{commuter} is ensured:
\beq \label{commuter101}
\ls Q_1, a^{\dagger}_{E_j}\rs \sim \text{O} (\epsilon). 
\eeq
A more rigorous approach to showing the existence of a suitable boundary observable can be found in Appendix \ref{sec:general_rs_proof}.

Let us now look at the second term in \eqref{commuter}. This renders $\braket{0, \mathcal{S}|\,a^{\dagger}_{E_j} Q_1|0, \mathcal{S}} = 0$, thereby satisfying the first condition in \eqref{refeeq}. Regarding the other condition in \eqref{refeeq}, using \eqref{commuter}, the second term gives us $\braket{0, \mathcal{S}|\, A_i \, a^{\dagger}_{E_j} Q_1|0, \mathcal{S}}$. Now consider the commutator
\beq \label{commuter1}
A_i \, a^{\dagger}_{E_j} = \ls A_i, a^{\dagger}_{E_j}\rs  +  a^{\dagger}_{E_j} A_i .
\eeq
Since $A_i$ is again an operator with energy much higher than $\delta$, the first term in \eqref{commuter1} is O$(\epsilon)$, and the second term is zero i.e. $\braket{0, \mathcal{S}|\,a^{\dagger}_{E_j} A_i \,  B_1|0, \mathcal{S}} = 0$. We can repeat the analysis for multi-particle states as well as straightforwardly generalize the result from $B_1$ to $B_i$. We obtain similar conclusions for multi-particle states, with products of creation operators replacing the single creation operator in the analog of \eqref{commuter} and \eqref{commuter1}.  Given the above discussed smearing conditions, operators in \eqref{exop2} represent valid operators $C_i$ satisfying constraints in \eqref{boundc}, with errors from smearing again giving rise to an O$\lc \epsilon\rc$ correction.

\subsubsection{Existence of $C_i$ using $\pdel$ and the boundary algebra}
In this subsection, we argue that we can always construct $C_i$ using $\pdel$ and other elements of the boundary algebra which satisfy constraints in \eqref{boundc}. This differs from our analysis in \S \ref{422} since our expressions for $C_i$ satisfying \eqref{boundc} are exact here, without any factors of O$(\epsilon)$.

 In general, to exactly construct $C_i$ satisfying the constraints \eqref{boundc}, we require other boundary operators along with the projector $\pdel$. We define $$\ket{B^{\perp}_i,\{s\}} \equiv \frac{\enc{1-\pdelta}}{\mathcal{N}_i}\ket{B_i,\{s\}}$$ 
where $\mathcal{N}_i$ is a normalization constant. As an example, we can read off $\mathcal{N}_i=\beta_i$ from \eqref{starwars}, when we work with the exact vacuum projector $\mathcal{P}_{0}$. On the lines of the construction of $C_i$ using $\mathcal{P}_{0}$, we write a candidate $C_i$, which is a sum of all possible outer products multiplied by undetermined coefficients  
\begin{equation} \label{kek}
\begin{split} 
    C_i &= \int \lc \prod_{l,m} ds_{l,m} \rc \sum_{E_j, E_k} \, \Theta(\delta - E_j) \, \Theta(\delta -E_k)\,  x^i_{j,k} \ket{E_j, \{s\}}\bra{E_k, \{s\}}\\
    &+ \int \lc \prod_{l,m} ds_{l,m} \rc \sum_{E_j} \, \Theta(\delta - E_j) \enc{y^i_j \ket{B^{\perp}_i, \{s\}} \bra{E_j,\{s\}} + \hc}\\
    &+ \int \lc \prod_{l,m} ds_{l,m} \rc z_i\ket{B^{\perp}_i, \{s\}} \bra{B^{\perp}_i, \{s\}}.
    \end{split}
\end{equation}
Note that here the elements $\ket{E_j, \{s\}}\bra{E_k, \{s\}}$ belong to the boundary algebra as argued in \cite{Laddha:2020kvp}. We will systematically fix some of the coefficients in \eqref{kek} as follows, where we will suppress corrections of O$(\sqrt{G_N})$ for presentation. The Hermiticity of $C_i$ implies $x^i_{j,k} = \lc x^i_{k,j} \rc^*$ and $z_i \in \mathbb{R}$. Imposing $C_i\ket{0,\{s\}} = B_i\ket{0,\{s\}} $, we fix the coefficients
\begin{equation}
    x^i_{m,0} = \int \lc \prod_{l,m} ds_{l,m} \rc \braket{E_m,\{s\}|B_i,\{s\}}, \qquad y^i_0 = \mathcal{N}_i.
\end{equation}
which ensures $\braket{A_j C_i} = \braket{A_j B_i}$. Given that we still have undetermined coefficients in $C_i$, we can always choose them in such a way that the absolute value of the largest eigenvalue is given by $\sqrt{\braket{B_i^2}}$, thereby giving us $\norm{C_i} = \braket{B_i^2} $.

As an example, we will demonstrate this for the case of the exact projector $\mathcal{P}_{0}$. Here we have $\mathcal{N}_i = \beta_i$. Consequently we obtain
$$\norm{C_i} = \frac{1}{2} \enc{\braket{B_i} + z_i + \sqrt{4\beta_i^2 + \enc{\braket{B_i}-z_i}^2}}.$$
Now requiring that the bound is satisfied, i.e. $\norm{C_i}^2 = \braket{B_i^2}$ gives us $z_i = -\braket{B_i}$, which again leads to the seemingly serendipitously constructed $C_i$ in \eqref{exop}. 

\subsection{The paradox and generalization to higher dimensions}
\label{messi}
In \S \ref{caccac}, using the wedge reconstruction, the boundary algebra, and the fact that the Hamiltonian in gravity is a boundary term, we have constructed operators living in the exterior region $C$ which essentially replicate the action of operators $B_i$ on the vacuum state, in three different fashions. Subsequently, we arrive at the following conclusion: 
\beq
\braket{C_{AC}} = \braket{C_{AB}} + \text{O} \lc \sqrt{G_N} \rc.
\eeq
Consequently at $x = \frac{1}{2}$, the correlator $\braket{C_{AC}} $ takes a maximum value $\braket{C_{AC}} = \frac{27\sqrt{2}}{16} + \text{O} \lc \sqrt{G_N} \rc + \text{O} \lc \epsilon \rc$.

After getting all our ingredients in place we will now pose the paradox in monogamy of entanglement. For the maximum violation at $x = \frac{1}{2}$, we obtain 
\beq \label{finoall}
\braket{C_{AB}}^2 + \braket{C_{AC}}^2 = 11.4 + \text{O} \lc \sqrt{G_N} \rc + \text{O} \lc \epsilon \rc > 8.
\eeq
Equation \eqref{finoall} contradicts the upper bound in \eqref{monent} and gives rise to the paradox in monogamy. As mentioned earlier, this is an O$(1)$ violation. The violation does not have a leading dependence on the IR cutoff $\delta$, an expected feature of a well-defined physical observable. 

An immediate generalization of the paradox in four dimensions is extending the same to general dimensions. In $d \neq 4$, the low energy vacuum structure of gravity is not concretely established (See \cite{Kapec:2015vwa, Hollands:2016oma, Aggarwal:2018ilg, Campiglia:2017xkp, He:2019pll, Campiglia:2016jdj, Sahoo:2018lxl} for recent discussions on the subject). Provided that the vacuum structure of gravity in higher dimensions has a similar form, i.e., there is a unique vacuum or degenerate vacua labelled by supertranslations, we can pose the paradox in precisely the same fashion we have done presently. Regarding additional symmetries, we can again treat them in a fashion similar to our treatment of supertranslations. 

We will point out why analogs of supertranslations in general dimensions are not in conflict with our calculation. The calculation of $\braket{C_{AB}}$ does not require us to go to the asymptotics since the operator insertions are deep inside the bulk, and hence our operator insertions do not change the supertranslation of the state on which they act. More precisely, these operator insertions are hard. The case of $\braket{C_{AC}}$ is a bit more subtle since it involves the construction of operators $Q_i$ and the projector $\pdel$ both of which have support near $\mathcal{I}^+_-$. However, from \eqref{qeye}, the action of $Q_i$ on the supertranslation fixed vacuum is precisely the action of the hard operators $B_i$ on the vacuum.  In addition, the projector $\pdel$ as defined in \eqref{pdelta} is diagonal in supertranslation labelled vacua. Consequently, the insertion of the operator $C_i$ within the vacuum to vacuum correlators does not introduce any new complications because of our construction and the very nature of the vacuum structure.  
 
\subsection{Resolution of the paradox}
\label{resopar}
In our calculation, we have explicitly pointed out small corrections of O$\lc \sqrt{G_N} \rc$ (See \eqref{finoall}). Hence the paradox cannot be resolved by introducing small corrections, as is the case for Hawking's original paradox and the bags of gold paradox \cite{Maldacena:2001kr, Barbon:2004ce, Anous:2016kss, Fitzpatrick:2016mjq, Chakravarty:2020wdm, Langhoff:2020jqa, Penington:2019kki}. Here the O$(1)$ violation indicates the existence of a severe flaw in our basic assumption, i.e., we have assumed that in the presence of gravity, our system admits a description in terms of a local quantum field theory. Building upon this assumption, we have factorized our Hilbert space into three different parts into three spatially disconnected and separated regions $A, B$, and $C$. 

However, it is a well-known fact that in gravity, the Hilbert space cannot be factorized due to the Gauss constraint. Consequently, our factorization into a tripartite system each described by a local QFT is incorrect, which resolves the paradox posed above. With gravity turned on, degrees of freedom in the region $B$ are secretly the same as degrees of freedom in the region $C$. Therefore it is incorrect to describe operators probing the underlying degrees of freedom using local quantum field theory, and we explicitly see an O$(1)$ violation if we assume a local quantum field theory setup in our case of empty flat space.

In a certain sense, we can observe this non-factorization of Hilbert space of the effective field theory based on spatial partitioning at the level of commutators itself \footnote{We thank Simon Caron-Huot for pointing out this issue.}. Note that since our operator insertions $A_i/B_i$ introduce energy into the bulk, the commutator $\ls H, B_i\rs \neq 0$. Since $C_i$ is a function of boundary projectors, in general the commutator $\ls C_j, B_i\rs \neq 0$. Following the Gauss constraint, this is a complementary objection to why we should not expect factorization based on spatial partitioning within a theory of gravity, even though effective field theory reasoning naively indicates otherwise. 

Since our calculation is performed in a general fashion, we can equivalently interchange the operators $A$ and $B$ describing the interior and the exterior respectively in \eqref{finoall} to set up an information-theoretic inequality again.  The resolution for this situation is precisely what black hole complementarity states, i.e., the interior operators are complicated polynomials of the exterior operators. In principle, our setup provides an explicit demonstration of the complementarity principle in flat space. Our boundary projector is a vital ingredient in this construction, allowing us to write down operators far away from the light cone that can probe the interior of the light cone. 

\section{Summary and Discussion}
\label{summarydiscussion}
We summarize the work, discuss some issues and write future directions here. Firstly, we introduced a formalism for treating Bell inequalities in a local QFT in flat space. We utilized the fact that monogamy of Bell correlations is a concrete measure for monogamy of entanglement and consequently used our formalism to compute Bell correlation between regions $A$ and $B$. Then bounded operators $C_i$ were constructed in a spacelike separated region $C$ close to the boundary using the Reeh Schlieder theorem and the boundary projector, which replicate the operators $B_i$'s action on the vacuum. Using this, a concrete paradox was posed in the monogamy of entanglement between the regions $A$, $B$ and $C$. We argued that the resolution to the paradox is as follows: in a theory of gravity, one cannot factorize the Hilbert space into subspaces describing spatially separated regions, which is necessary to set up a paradox in the monogamy of entanglement.

As we discussed, in canonical gravity, the Hamiltonian is a boundary term that plays a crucial role in constructing bounded operators in the region $C$ that replicate operators $B_i$. The fact that the Hamiltonian is a boundary term is an essential feature of gravity, which strongly hints at non-local aspects inbuilt within theories of gravity\footnote{In a certain sense, this is a generalization of the Gauss law in Newtonian gravity, where the effect of any massive insertion at a point within a spacelike separated Gaussian surface is necessarily manifested on the surface.}. Note that this feature is unique to gravity and is not true of other theories, say theories with Gauss constraints. Case in point, operator insertions with zero charge in gauge theories do not affect the field strength residing on the Gaussian surface. In gravity, an operator insertion necessarily changes the stress-energy tensor, and consequently, one cannot introduce invisible operator insertions\footnote{This argument is also tied to why we cannot write diffeomorphism invariant local observables in gravity.}.

We now discuss the relation of our model to the monogamy paradox for old black holes in flat space. Since non-local effects of gravity play a primary role in our problem, it is only natural to assume that such effects play a similar role in the black hole problem\footnote{In fact, such non-local effects are a generic feature of quantum gravity, and manifestly reveal them under extreme situations  \cite{Ghosh:2017pel}.}. The operators $C_i$ in our problem are in a spirit similar to complicated operators situated far away from a black hole used to extract information from Hawking radiation. Our construction also emphasizes the usage of CHSH correlations in studying the monogamy of entanglement paradox, primarily how CHSH correlations can be used to quantify entanglement. The study of these correlators is necessary since standard measures of entanglement like Von Neumann entropy are not well defined in gravity.

Note that the monogamy paradox is conventionally posed within the context of old black holes. However, our discussion only relies upon the entanglement of modes across the horizon and the boundary. Consequently we do not require an old black hole to pose the paradox, which is reflected in the fact that the validity of operators $C_i$ does not involve any particular time scale. In line with the principle of holography of information \cite{Laddha:2020kvp, Raju:2020smc, Chowdhury:2020hse}, this is because the information about the non-boundary regions is \textit{always contained} within $\mathcal{I}^+_-$. 

The issue with writing down a similar construction for evaporating black holes is that we require a projector onto the space of black hole microstates, which we presently do not understand how to construct. Consequently, it is not easy to write bounded operators in a region far away from the black hole, which can be used to write down CHSH correlations. However, there is no problem with calculating $C_{AB}$ correlator between modes just inside and outside the horizon. Formulating the paradox in our toy model's fashion also shows that we do not need any modified structure in the black hole interior, as is the case with firewall and fuzzball constructions. Instead, such a paradox in monogamy is a natural consequence of wrongly treating gravity as a local quantum field theory.

Before we conclude, we list out some related open questions. In our case, we need to go very close to future null infinity to construct a projector onto our ground state. In line with our holographic intuition that gravity knows about quantum information inside a given region, is it possible to construct a similarly approximate projector onto the vacuum at a finite radius? Finding such a projector will be pretty valuable not only as an independent problem for our flat space toy model but also to pose a similar resolution of the monogamy paradox for dS black holes. Besides, such a projector will be pretty valuable for understanding aspects of the principle of holography of information for compact spacetimes, where naively a projector will project onto all physical states in the Hilbert space since there is no boundary, and consequently, we need a projector at a finite radius. Another problem is to write down the projector onto the space of all black hole microstates in flat space and AdS, which will allow us to write down a more accurate toy model. A distant direction is to understand the asymptotic vacuum structure in general dimensions, which will be helpful to pose the toy model concretely in such dimensions. We envisage our present work as a starting point to address some of these issues in the near future.

 \begin{acknowledgments}
We thank Suvrat Raju for suggesting the problem and guidance regarding this work. We also thank Simon Caron-Huot for raising important points regarding the initial draft.  We are also grateful to Chandramouli Chowdhury, Victor Godet, Diksha Jain, Alok Laddha, R. Loganayagam, Ruchira Mishra, Olga Papadoulaki, Siddharth G. Prabhu, Omkar Shetye,  Pushkal Shrivastava, and Akhil Sivakumar for helpful discussions regarding the work. We acknowledge support of the Department of Atomic Energy, Government of India, under project number RTI4001. The authors also acknowledge gratitude to the people of India for their steady and generous support to research in basic sciences.

\end{acknowledgments}
\appendix 

\section{Projectors onto smeared modes' vacua}
\label{p0construction}
In this section we shall verify the expression for projector onto vacuum (\ref{p0}). We first take a variable transformation, $z = t_1+ i t_2$ and $z^*= t_1-it_2$. With
${\alpha}_s=\tfrac{1}{\sqrt{2}}({X}_s+ i {\Pi}_s)$ and using Baker–Campbell–Hausdorff lemma, we can write the projector as,
\begin{equation*}
    P_s= - \dfrac{1}{\pi^2}\int d^2z\int_{0}^{2 \pi}d \theta_s \frac{e^{-z \bar{z}(1-i \tan\theta_s)}} {e^{i \theta_s }-1 - \epsilon}  e^{- \beta(\theta_s) \bar{z} {\alpha}_s^{\dagger}} e^{- \beta(\theta_s)z {\alpha}_s  }
\end{equation*}
where $\beta(\theta) = \sqrt{2i\tan \theta}$. Let us calculate $\langle i_s | P_s | j_s \rangle $, where $| i_s \rangle , | j_s \rangle$ are number states corresponding to oscillator labelled by $s$. We get,
\begin{equation}
    \langle i_s | e^{- \beta(\theta) \bar{z} {\alpha}_s^{\dagger}} e^{- \beta(\theta)z {\alpha} _s}  | j_s \rangle 
         =    \sum_{m =0 }^{\infty} \sum_{n= 0}^{\infty} (-\beta(\theta_s))^{m+n}\dfrac{\bar{z}^m}{m!} \dfrac{{z}^{n}}{n!} \langle i_s | {\alpha}_s^{ \dagger \; m} {\alpha}_s^{n} | j_s \rangle 
\end{equation}
This is only non-zero if $ n +i_s= m+j_s $. If we also perform the $z,\bar{z}$ integral with $z=r e^{i \phi}$ and  $\bar{z}=r e^{-i \phi}$, that further constrains us with a $\delta_{i_s,j_s}$ factor. Hence,
\begin{equation}
     \langle i_s | P_s | j_s \rangle   = - \frac{1}{\pi} \int_{0}^{2 \pi}d \theta_s \sum _{n = 0}^{\infty} \frac{(2 i \tan\theta_s)^{n}} {(1- i \tan\theta_s)^{n+1} (e^{i \theta_s }-1-\epsilon)}  \bigg[\dfrac{\delta_{i_s,j_s}}{n!} \langle i_s |  {\alpha}_s^{\dagger \; m} {\alpha}_s^{ n} | j_s \rangle  \bigg]_{m = n} 
\end{equation}
The term inside third braces is,
\begin{equation}
\delta_{i_s,j_s}\dfrac{1}{n!}\langle i_s| {\alpha}_s^{\dagger n }\alpha_s ^{n} | j_s \rangle = \begin{cases} \delta_{i_s,j_s} \dfrac{i_s!}{n ! (n-i_s)!}  \; \; \;  &\text{for} \; \; \;  n \leq i_s,\\
0 \; \; \; &\text{for} \; \; \;  n > i_s.\\
\end{cases}
\end{equation}
Summing over $n$ and further changing variable $\omega = e^{i \theta_s}$, we have a contour integral
\begin{equation}
    \langle i_s | P_s | j_s \rangle = -\dfrac{\delta_{i_s,j_s}}{2 \pi i } \oint_{|\omega|= 1} d \omega \dfrac{\omega^{2 i _s -1}(\omega^2+1)}{(\omega-1-\epsilon)}.
\end{equation}
\begin{figure}
    \centering
    \includegraphics[scale=0.5]{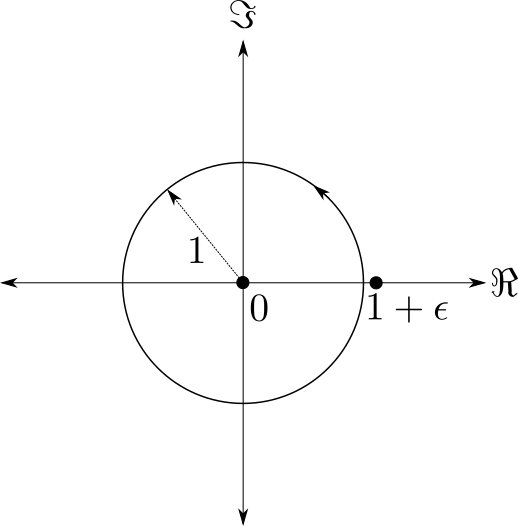}
    \caption{Contour used in the smeared vacuum projector calculation. There exist two poles at $0$ and $1+ \epsilon$, amongst which only the former contributes to the unit circle contour integral.}
    \label{fig:contour2}
\end{figure}
With $\epsilon > 0 $ as shown in \cref{fig:contour2} the contour evaluates to
\begin{equation}
    \langle i_s | P_s | j_s \rangle = \delta _{i_s,j_s} \delta_{i_s,0} 
\end{equation}
Hence,
\begin{equation}
     P_s = - \frac{1}{\pi^2}\int d t_1 dt_2 \int_{0}^{2 \pi} d \theta_s \dfrac{1}{e^{i \theta_s}-1- \epsilon} e^{- (t_1^2+ t_2^2) - \kappa(\theta_s)(t_1 {X}_s- t_2 {\Pi}_s)} = | 0_s\rangle \langle 0_s|
\end{equation}
\section{Explicit commutator of smeared Rindler modes}
\label{appb}
Consider the commutator of the modes on region $A$ first. The modes are given by:
\begin{align}
    \alpha_A &= \frac{1}{\sqrt{V_\Omega}} \int \frac{dU}{U} \int d^{d-2}\Omega \, r_A^{\frac{(d-2)}{2}} \enc{\frac{U}{U_0}}^{i\omega_0} \T (U) \, \phi(t_A(U),r_A(U),\Omega),\\
     \alpha^{\dagger}_A&= \frac{1}{\sqrt{V_\Omega}} \int \frac{dU}{U} \int d^{d-2}\Omega \, r_A^{\frac{(d-2)}{2}} \enc{\frac{U}{U_0}}^{-i\omega_0} \T^*(U) \, \phi(t_A(U),r_A(U),\Omega)
\end{align}
We integrate $\alpha_A^\dagger$ by parts, and the only part of $\alpha^\dagger_A$ that contributes to the commutator $\ls \alpha_A, \alpha_A^\dagger \rs$ is 
\begin{equation}
       \frac{1}{\sqrt{V_\Omega}}\int d^{d-2}\Omega \int dU \, r_A^{\frac{(d-2)}{2}} \partial_U \phi(U,-2v_0,\Omega) \int \frac{d\nu}{i\nu} \tilde{\T}^{*}(\nu) \enc{\frac{U}{U_0}}^{-i\nu} 
\end{equation}
Note that here we have used $\T(U)\enc{\frac{U}{U_0}}^{i\omega_0} = \int d\nu \, \Tilde{\T}(\nu) \enc{\frac{U}{U_0}}^{i\nu}$ to perform the following replacement
 \begin{equation}
     \int_0^{U} \frac{dU'}{U'} \T(U')\enc{\frac{U'}{U_0}}^{i\omega_0} = \int \frac{d\nu}{i\nu} \tilde{\T}(\nu) \enc{\frac{U}{U_0}}^{i\nu}.
 \end{equation}
In terms of the light cone coordinates, the annihilation operator is given by
\begin{equation}
    \alpha_A = \frac{1}{\sqrt{V_\Omega}} \int  \frac{dU}{U}\int d^{d-2}\Omega \, r_A^{\frac{(d-2)}{2}} \phi(U,-2v_0,\Omega) \int d\nu \tilde{\T}(\nu) \enc{\frac{U}{U_0}}^{i\nu}
\end{equation}
Using the null surface canonical commutation \cite{suzuki_null_quant} relation 
\begin{equation}
    [\phi(U_1,V,\Omega_1), \partial_U \phi(U_2,V,\Omega_2)] = \frac{i\delta(U_1 - U_2) \delta(\Omega_1,\Omega_2)}{2r_1^{d-2}}
\end{equation}
we get
\begin{align}
    [\alpha_A, \alpha_A^{\dagger}] &= \frac{1}{2} \int \frac{d\Omega}{V_\Omega} \int \frac{d\nu_1 d\nu_2}{\nu_2} \, \tilde{\T}(\nu_1) \tilde{\T}^* (\nu_2) \int \frac{dU}{U} \enc{\frac{U}{U_0}}^{i(\nu_1 - \nu_2)} \\
    &=  \frac{1}{2} \int \frac{d\Omega}{V_\Omega} \int \frac{d\nu_1 d\nu_2}{\nu_2} \, \tilde{\T}(\nu_1) \tilde{\T}^* (\nu_2)\,  2\pi \delta(\nu_1 - \nu_2) \\
    &= 1
\end{align}
where we have used the normalization $\int \frac{d\nu}{\nu} |\tilde{\T}(\nu)|^2 = \frac{1}{\pi}$.

\section{Computation of $\braket{\mathcal{G}}$}
\label{app:G}
We will discretize the frequency space with a $\Delta$ gap for ease of calculation. Towards the end of this appendix, we will go back to the continuous limit by taking $\Delta\rightarrow 0$. Performing the discretization, the global mode commutation relation (\ref{cancomm}) becomes
\begin{equation}
    [{a}_{n,l},{a}_{n',l'}^{\dagger}] = \frac{\delta_{n,n'}\delta_{l,l'}}{\Delta},
\end{equation}
where we have labelled the frequency $\omega$ with integer $n$. The global mode decomposition (\ref{bogo}) now looks like ${\alpha}_s= \Delta\sum_{n,l} h_{s}(n,l) {a}_{n,l} +  g_{s}^{*}(n,l) {a}_{n,l}^{\dagger}$. Using the BCH lemma, we can decompose the $B$ piece in $\braket{\mathcal{G}}$ in terms of creation and annihilation operators as 
\begin{equation}
    e^{v_2 {\alpha}^{\dagger}_B} e^{ (\Tilde{y}_1 {X}_B- \Tilde{y}_2 {\Pi}_B) } e^{\zeta_2 {\alpha}_B} = e^{ \big(v_2 +  \tfrac{(\Tilde{y}_1- i \Tilde{y}_2)}{\sqrt{2}} \big) {\alpha}_B ^{\dagger} +  \big( \zeta_2 + \tfrac{(\Tilde{y}_1+ i \Tilde{y}_2)}{\sqrt{2}} \big) {\alpha}_B }  e ^{- \tfrac{1}{2} \big( v_2 \tfrac{(\Tilde{y}_1+ i \Tilde{y}_2)}{\sqrt{2}} + \zeta_2  \tfrac{(\Tilde{y}_1- i \Tilde{y}_2)}{\sqrt{2}} + v_2 \zeta_2\big)}.
\end{equation}
Similarly decomposing the $A$ piece and then writing both in terms of global modes, we get 
\begin{equation}
    \langle \mathcal{G} \rangle = \frac{\bigg\langle \exp\enc{\Delta\sum\limits_{n,l} \big( u_B (n,l)+ u_A(n,l) \big) {a}_{n,l} + \big( u_B'(n,l) + u_A '(n,l)\big) {a}_{n,l}^{\dagger}} \bigg\rangle}{e ^{ \tfrac{1}{2} \big( v_2 \tfrac{(\Tilde{y}_1+ i \Tilde{y}_2)}{\sqrt{2}} + \zeta_2  \tfrac{(\Tilde{y}_1- i \Tilde{y}_2)}{\sqrt{2}} + v_2 \zeta_2\big)}  e ^{ \tfrac{1}{2} \big( v_1 \tfrac{(\Tilde{t}_1+ i \Tilde{t}_2)}{\sqrt{2}} + \zeta_1  \tfrac{(\Tilde{t}_1- i \Tilde{t}_2)}{\sqrt{2}} + v_1 \zeta_1\big)}},
    \label{eq:g_firstform}
\end{equation}
where 
\begin{align}
    u_{A}(n,l) &= \big(v_1 +  \tfrac{(\Tilde{t}_1- i \Tilde{t}_2)}{\sqrt{2}} \big) g_{A}(n,l)  + \big( \zeta_1 + \tfrac{(\Tilde{t}_1+ i \Tilde{t}_2)}{\sqrt{2}}  \big) h_{A}(n,l), \\
    u_ {A}'(n,l) &= \big(v_1 +  \tfrac{(\Tilde{t}_1- i \Tilde{t}_2)}{\sqrt{2}} \big) h_{A}^*(n,l)  + \big( \zeta_1 + \tfrac{(\Tilde{t}_1+ i \Tilde{t}_2)}{\sqrt{2}}  \big) g_{A}^*(n,l), \\
    u_{B}(n,l) &= \big(v_2 +  \tfrac{(\Tilde{y}_1- i \Tilde{y}_2)}{\sqrt{2}} \big) g_{B}(n,l)  + \big( \zeta_2 + \tfrac{(\Tilde{y}_1+ i \Tilde{y}_2)}{\sqrt{2}}  \big) h_{B}(n,l), \\ 
    u_ {B}'(n,l) &=   \big(v_2 +  \tfrac{(\Tilde{y}_1- i \Tilde{y}_2)}{\sqrt{2}} \big) h_{B}^*(n,l)  + \big( \zeta_2 + \tfrac{(\Tilde{y}_1+ i \Tilde{y}_2)}{\sqrt{2}}  \big) g_{B}^*(n,l).
\end{align}
Next we use a simple result involving coherent states of harmonic oscillators to simplify our expressions further. Consider a system of oscillators with ground states $\ket{0_i}$, where ($i=1,2,...,\infty$), with commutation relations $[\hat{\alpha}_i,\hat{\alpha}^\dagger_j]=\delta_{ij}$. Define the combined ground state of the system as $\ket{0}\equiv\bigotimes_i \ket{0_i}$. This setup is intended to mimic the global modes $a_{n,l}$, as in our theory the global vacuum is indeed the tensor product of all the different global mode vacua. A coherent state in the $j$'th oscillator is given by $\ket{z_j}\equiv e^{z_j \hat{\alpha}_j^{\dagger}}\ket{0_j}$, and the inner product between two such states is $\braket{z_j|z'_j} = e^{z^*_j z'_j}$ \cite{book:brown_qft}. Then we have
\begin{align}
    \big\langle 0 \big| e^{\sum\limits_i z^*_i \hat{\alpha}_i + z_i' \hat{\alpha}_i^{\dagger} }   \big| 0 \big\rangle &= e^{-\frac{1}{2}\sum_{i} z^*_i z'_i} \big\langle 0 \big| e^{\sum\limits_i z^*_i \hat{\alpha}_i }  e^{\sum\limits_j z_j' \hat{\alpha}_j^{\dagger} }   \big| 0 \big\rangle \\
    &= e^{-\frac{1}{2}\sum_{i} z^*_i z'_i} \prod\limits_{ij} \big\langle 0_i \big|  e^{ z^*_i \hat{\alpha}_i } e^{ z_j' \hat{\alpha}_j^{\dagger} }   \big| 0_j \big\rangle \\
    &= e^{-\frac{1}{2}\sum_{i} z^*_i z'_i} \prod\limits_{j} \braket{ z_j | z'_j } \\
    &= e^{\frac{1}{2}\sum\limits_{i} z^*_i z_i'}.
\end{align}
To make use of this in simplifying (\ref{eq:g_firstform}), we identify $\sqrt{\Delta}a_{n,l}$ with $\hat{\alpha}_j$. This gives us terms like $\sum_{n}\Delta u_A(n,l)u'_B(n,l)$ on top of the exponential. Taking the limit $\Delta\rightarrow 0$, the sum $\sum_j$ goes to an intgral and the whole expression 
simplifies to
\begin{equation}
    \langle \mathcal{G} \rangle = \frac{ \exp\enc{\frac{1}{2}  \enc{ u_B + u_A}\cdot \enc{ u_B' + u_A '}}}{e ^{ \tfrac{1}{2} \big( v_2 \tfrac{(\Tilde{y}_1+ i \Tilde{y}_2)}{\sqrt{2}} + \zeta_2  \tfrac{(\Tilde{y}_1- i \Tilde{y}_2)}{\sqrt{2}} + v_2 \zeta_2\big)}  e ^{ \tfrac{1}{2} \big( v_1 \tfrac{(\Tilde{t}_1+ i \Tilde{t}_2)}{\sqrt{2}} + \zeta_1  \tfrac{(\Tilde{t}_1- i \Tilde{t}_2)}{\sqrt{2}} + v_1 \zeta_1\big)}},
    \label{eq:g_secondform}
\end{equation}
where $u_A \cdot u_B \equiv \sum_{l}\int d\omega\, u_A(\omega,l) u_B (\omega, l)$. Using the $f_p$ defined in (\ref{eq:fm_defs}), we obtain
\begin{align}
    u_A+ u_B &= \frac{1}{\sqrt{2}}[f_1 (\Tilde{t}_1 + \zeta^+_1   ) + f_2 (-\Tilde{t}_2+ i \zeta_1^-  ) + f_3(\Tilde{y}_1 + \zeta^+_2   ) + f_4 (-\Tilde{y}_2+ i \zeta_2^-  ) ]\\
    u'_A + u'_B &= \frac{1}{\sqrt{2}}[f_1^* (\Tilde{t}_1 + \zeta^+_1   ) + f_2^* (-\Tilde{t}_2+ i \zeta_1^-  ) + f_3^*(\Tilde{y}_1 + \zeta^+_2   ) + f_4^* (-\Tilde{y}_2+ i \zeta_2^-  ) ].
\end{align}
Re-arranging the terms to gather the $f_p$'s together and using the $m_q$ defined in (\ref{eq:fm_defs}), we finally obtain
\begin{equation}
    \langle \mathcal{G} \rangle = \exp \enc{ \frac{1}{8}\sum\limits_{p,q=1}^{4}  (f _p \cdot f_q^* + f_q \cdot f_p^*)  m_p m_q - \frac{\mathcal{R}}{2}}.
\end{equation}

\section{Bogoliubov coefficients and $\braket{C_{AB}} \geq 2$}
In this appendix, we demonstrate the calculation of the Bogoliubov coefficients and show that $\braket{C_{AB}} \geq 2$.

\subsection*{Bogoliubov coefficients of local Rindler-to-global modes}
\label{bogorindmink}
From \eqref{lfl} we can read off the Bogoliubov coefficients using the large frequency limit, which are given by
\beq
\begin{split}
    h_A(\omega,0) &= \frac{1}{\sqrt{\pi \omega}} \int \frac{dU}{U} \enc{\frac{U}{U_0}}^{i\omega_0} \T(U) e^{-i\omega t_A} \cos\enc{\omega r_A - \frac{(d-2)\pi}{4}},\\
    g^*_A(\omega,0) &= \frac{1}{\sqrt{\pi \omega}} \int \frac{dU}{U} \enc{\frac{U}{U_0}}^{i\omega_0} \T(U) e^{i\omega t_A} \cos\enc{\omega r_A - \frac{(d-2)\pi}{4}},\\
    h_B(\omega,0) &= \frac{1}{\sqrt{\pi \omega}} \int \frac{dU}{U} \enc{\frac{U}{U_0}}^{-i\omega_0} \T(U) e^{-i\omega t_B} \cos\enc{\omega r_B - \frac{(d-2)\pi}{4}},\\
    g^*_B(\omega,0) &= \frac{1}{\sqrt{\pi \omega}} \int \frac{dU}{U} \enc{\frac{U}{U_0}}^{-i\omega_0} \T(U) e^{i\omega t_B} \cos\enc{\omega r_B - \frac{(d-2)\pi}{4}}.  
\end{split}
\eeq
The above Bogoliubov coefficients are written in terms of integrals over $U$. We can perform these integrals using our conditions on the tuning function in \eqref{tuning1} and \eqref{tuva}. Since the form of the integrals is similar, we will demonstrate this by evaluating the $h_A(\omega,0)$ integral:
\begin{equation*}
\begin{split}
    h_A(\omega,0) &= \frac{1}{\sqrt{\pi \omega}} \int \frac{dU}{U} \enc{\frac{U}{U_0}}^{i\omega_0} \T(U) e^{-i\omega t_A} \cos\enc{\omega r_A - \frac{(d-2)\pi}{4}} \\
    &= \frac{1}{2\sqrt{\pi\omega}} \int \frac{dU}{U} \enc{\frac{U}{U_0}}^{i\omega_0} \T(U) \encsq{e^{-i\xi_1} e^{-i\omega U} + e^{i\xi_2}} \\
    &= \frac{1}{2\sqrt{\pi\omega}} \int d\nu \frac{\tilde{\T}(\nu)}{(U_0)^{i\nu}} \encsq{e^{-i\xi_1} \int dU \, U^{i\nu-1} e^{-i\omega U} + e^{i\xi_2} \int \frac{dU}{U} U^{i\nu}} \\
    &= \frac{e^{-i\xi_1}}{2\sqrt{\pi\omega}} \int d\nu \frac{\tilde{\T}(\nu)}{(\omega U_0)^{i\nu}} \int dx\, x^{i\nu-1} e^{-i x} + 0
\end{split}
\end{equation*}
\begin{figure}
    \centering
    \includegraphics[scale=0.5]{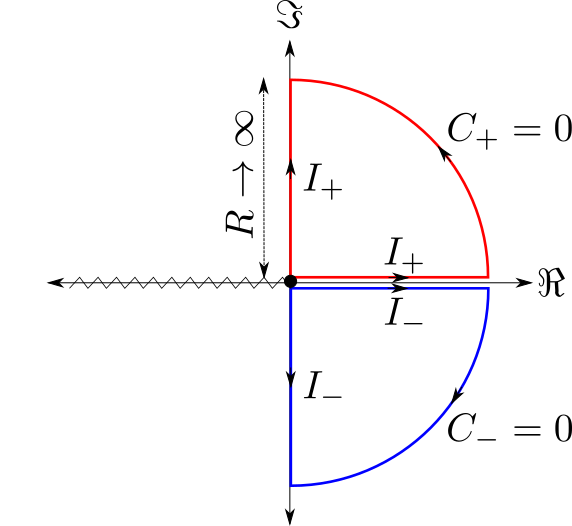}
    \caption{The red and the blue contours are respectively used for the integrals $I_{\pm}=\int dx\, e^{\pm ix} x^{i\nu-1}$. Both the curved contours $C_{\pm}$ give $0$ in the limit $R\rightarrow\infty$. Since there are no poles inside either contour (they are slightly separated from the real axis), the total contours also give 0. This lets us equate the integrals over the real and the imaginary axes for both the $e^{ix}$ and $e^{-ix}$ cases. We keep the branch cut (occurring due to $x^{i\nu}$) on the negative real axis so it doesn't interfere with the calculation.}
    \label{fig:contour1}
\end{figure}
where $\xi_1 = \frac{(d-2)\pi}{4} - \omega r_0$, $\xi_2 = \frac{(d-2)\pi}{4} - \omega (r_0 - 2v_0)$. The second term in the third line vanishes because $\int \frac{dU}{U} \enc{\frac{U}{U_0}}^{i\nu} = 2\pi \delta(\nu)$, and using \eqref{tuva} $\tilde{\T}(\nu)$ vanishes at $\nu=0$. The $x$ integral can be evaluated by choosing a contour shown in \cref{fig:contour1}. We encounter such $x$ integrals in the expressions for the other Bogoliubov coefficients as well, where we similarly choose appropriate contours and obtain the following values for the integrals
\begin{equation}
    \int_0^\infty dx \, x^{i\nu -1} e^{\pm i x} = e^{\mp \pi \nu /2} \, \Gamma (i\nu).
\end{equation}
Thus the Bogoliubov coefficients can be conveniently summarized as
\beq \label{bogofin0}
\begin{split}
    h_A (\omega,0) &= \frac{e^{-i \xi_1}}{2\sqrt{\pi\omega}} \int d\nu\, e^{\pi \nu/2} (\omega U_0)^{-i\nu} \Gamma(i\nu) \tilde{\T}(\nu)  , \\
    g_A^* (\omega,0) &= \frac{e^{i \xi_1}}{2\sqrt{\pi\omega}} \int d\nu\, e^{-\pi \nu/2} (\omega U_0)^{-i\nu} \Gamma(i\nu) \tilde{\T}(\nu)  , \\
    h_B (\omega,0) &= \frac{e^{-i \xi_1}}{2\sqrt{\pi\omega}} \int d\nu\, e^{\pi \nu/2} (\omega U_0)^{i\nu} \Gamma(-i\nu) \tilde{\T}^*(\nu)  ,\\ 
    g_B^* (\omega,0) &= \frac{e^{i \xi_1}}{2\sqrt{\pi\omega}} \int d\nu\, e^{-\pi\nu/2} (\omega U_0)^{i\nu} \Gamma(-i\nu) \tilde{\T}^{*}(\nu).
\end{split}
\eeq
\subsection*{Calculation of $\braket{C_{AB}} \geq 2$ for entangled Rindler modes}
\label{matrixapp}
Here we will demonstrate the calculation of the element $f_1 \cdot f_1^{*}=h_A^*\cdot h_A + g_A^*\cdot g_A + h_A^* \cdot g_A + g_A^* \cdot h_A$. The typical integral encountered here is of the form
\begin{equation*}
\begin{split}
    h_A \cdot h_A^* &=  \int \frac{d\omega}{4\pi\omega} \int d\nu_1 d\nu_2\, e^{\pi \enc{\nu_1 + \nu_2}/2} \enc{\omega U_0}^{i\enc{\nu_2-\nu_1}} \Gamma (i\nu_1) \Gamma^* (i\nu_2) \tilde{\T}(\nu_1) \tilde{\T}^*(\nu_2) \\
    &= \frac{1}{2} \int d\nu_1 d\nu_2\, e^{\pi \enc{\nu_1 + \nu_2}/2} \, \Gamma (i\nu_1) \Gamma^* (i\nu_2) \tilde{\T}(\nu_1) \tilde{\T}^*(\nu_2) \int \frac{d\omega}{2\pi \omega} \enc{\omega U_0}^{i\enc{\nu_2-\nu_1}}\\
    &= \frac{1}{2} \int d\nu_1 d\nu_2\, e^{\pi \enc{\nu_1 + \nu_2}/2} \, \Gamma (i\nu_1) \Gamma^* (i\nu_2) \tilde{\T}(\nu_1) \tilde{\T}^*(\nu_2) \delta (\nu_1 -\nu_2) \\
    &= \frac{1}{2} \int d\nu\,  e^{\pi \nu}\,  |\Gamma (i\nu)|^2 | \, \tilde{\T}(\nu)|^2\\
    &= \frac{\pi}{2} \int d\nu\, \frac{e^{\pi\nu}}{\nu \sinh (\pi\nu)} |\tilde{\T}(\nu)|^2\\
    &= \frac{e^{\pi \omega_0}}{2\sinh (\pi\omega_0)}\equiv  \frac{1}{1-x^2}.\\
\end{split}
\end{equation*}
where $x\equiv e^{-\pi\omega_0}$. Here in the fifth step, we have used the identity $|\Gamma(i\nu)|^2 = \frac{\pi}{\nu \sinh (\pi\nu)}$ and used the fact that $|\tilde{\T}(\nu)|^2/\nu$ is sharply peaked around $\omega_0$ to go the sixth step. We will now show that $h_A \cdot g_A^*$ is zero.
\begin{equation*}
\begin{split}
    h_A \cdot g_A^* &= \int \frac{d\omega}{4\pi\omega} \int d\nu_1 d\nu_2\, e^{\pi \enc{\nu_1 - \nu_2}/2} \enc{\omega U_0}^{-i\enc{\nu_2+\nu_1}} \Gamma (i\nu_1) \Gamma^* (i\nu_2) \tilde{\T}(\nu_1) \tilde{\T}^*(\nu_2) \\
    &= \frac{1}{2} \int d\nu_1 d\nu_2\, e^{\pi \enc{\nu_1 - \nu_2}/2} \Gamma (i\nu_1) \Gamma^* (i\nu_2) \tilde{\T}(\nu_1) \tilde{\T}^*(\nu_2) \delta (\nu_1 +\nu_2) \\
    &= \frac{1}{2} \int d\nu\,  e^{\pi \nu} \Gamma (i\nu) \Gamma^* (-i\nu) \tilde{\T}(\nu) \tilde{\T}^*(-\nu)\\
    &= 0
\end{split}
\end{equation*}
The final step follows due to the fact that within the integral $\int d\nu \tilde{\T}(\nu) \tilde{\T}^*(-\nu)$, when $\tilde{\T}(\nu)$ peaks at $\nu=\omega_0$, the other term goes to zero, i.e. $\tilde{\T}^*(-\nu) = \tilde{\T}^*(-\omega_0) \approx 0$. 
The rest of the terms are evaluated by straightforward replication of the above logic. We similarly evaluate the following expressions in order to completely determine the $ff$ matrix.
\begin{equation} \label{ff0}
\begin{split}
    h_A \cdot h_A^* &= h_B \cdot h_B^* = \frac{1}{1-x^2}; \\
    g_A \cdot g_A^* &= g_B \cdot g_B^*  = \frac{x^2}{1-x^2}; \\
    h_A \cdot g_B^* &= g_A \cdot h_B^* = \frac{x}{1-x^2};\\
    h_A \cdot g_A^* &= h_A \cdot h_B^* = g_A \cdot g_B^* = h_B \cdot g_B^* =  0. \\
\end{split}
\end{equation}
Substituting the expressions in \eqref{ff0} in $f_p \cdot f_q^* + f_p^* \cdot f_q$, we obtain
\begin{equation} \label{ff2}
    f_p \cdot f_q^* + f_q \cdot f_p^* = \frac{2}{1-x^2} \begin{pmatrix}
    1+x^2 & 0 & 2x & 0 \\
    0 & 1+x^2 & 0 & -2x\\
    2x & 0 & 1+x^2 & 0\\
    0 & -2x & 0 & 1+x^2
    \end{pmatrix}.
\end{equation}

\section{Proof of $\braket{A_j C_i} = \braket{A_j B_i} + \text{O} \lc \sqrt{G_N} \rc$ and boundedness of $C_i$ of \S \ref{cpzero}}
\label{appc}
In this appendix, we will show that the operators $C_i$ constructed as
\begin{equation}
    C_i \equiv \frac{\braket{B_i^2}\enc{Q_i\pvacex+ \pvacex Q_i^\dagger- \braket{B_i}\pvacex} - \braket{B_i}Q_i \pvacex Q_i^\dagger}{\braket{B_i^2}-\braket{B_i}^2}.
\end{equation}
do indeed mimic the contribution of operators $B_i$ in the two-point correlators and are bounded.

\subsection*{Proof of $\braket{A_jC_i}=\braket{A_jB_i}$}
First we note that by construction we have $Q_i \ket{0,\{s\}} = B_i \ket{0,\{s\}}$ for all sectors $\{s\}$, which guarantees $\mathcal{P}_0 Q^\dagger_i = \mathcal{P}_0 B_i^\dagger = \mathcal{P}_0 B_i$. Further, the fact that $B_i$ is block diagonal in and independent of supertranslation sectors, allows us to write $\braket{0, \{s\} | B_i|0,\{s'\}} = K \delta(\{s\}-\{s'\})$. Taking the expectation value w.r.t. the smeared vacuum $\ket{0,\mathcal{S}}$ gives us $K = \braket{0,\mathcal{S}|B_i|0,\mathcal{S}}$. Thus we have
\begin{equation}
    \braket{0, \{s'\} | B_i|0,\{s\}} = \braket{0,\mathcal{S}|B_i|0,\mathcal{S}} \delta (\{s'\}-\{s\}).
\end{equation}
So, we have
\begin{equation}
    \begin{split}
        \mathcal{P}_0 Q_i^\dagger \ket{0,\mathcal{S}} &=  \mathcal{P}_0 B_i \ket{0,\mathcal{S}}\\
        &= \int \lc \prod_{l,m} ds_{l,m} \rc \int \lc \prod_{l,m} ds'_{l,m} \rc\, \ket{0,\{s'\}} \braket{0,\{s'\}|B_i|0,\{s\}} \mathcal{S}\enc{\{s\}}  + \oG\\
        &= \int \lc \prod_{l,m} ds_{l,m} \rc \int \lc \prod_{l,m} ds'_{l,m} \rc\, \ket{0,\{s'\}} \braket{B_i} \delta(\{s\}-\{s'\}) \mathcal{S}\enc{\{s\}} +\oG\\
        &= \braket{B_i} \int \lc \prod_{l,m} ds_{l,m} \rc\, \mathcal{S}\enc{\{s\}} \ket{0,\{s'\}}+ \oG\\
        &= \braket{B_i} \ket{0,\mathcal{S}} + \oG.
    \end{split}
\end{equation}
Now
\begin{equation}
    \begin{split}
    C_i\ket{0,\mathcal{S}} &= \frac{\braket{B_i^2}\enc{Q_i\ket{0,\mathcal{S}}+ \pvacex Q_i^\dagger\ket{0,\mathcal{S}}- \braket{B_i} \ket{0,\mathcal{S}}} - \braket{B_i} Q_i\pvacex Q_i^\dagger \ket{0,\mathcal{S}}}{\braket{B_i^2}-\braket{B_i}^2} +\text{O} \lc \sqrt{G_N} \rc\\
    &= \frac{\braket{B_i^2}\enc{Q_i\ket{0,\mathcal{S}}+ \braket{B_i} \ket{0,\mathcal{S}} - \braket{B_i} \ket{0,\mathcal{S}}} - \braket{B_i} \braket{B_i} Q_i \ket{0,\mathcal{S}}}{\braket{B_i^2}-\braket{B_i}^2}+\text{O} \lc \sqrt{G_N} \rc\\
    &= \frac{\enc{\braket{B_i^2}-\braket{B_i}^2}Q_i\ket{0,\mathcal{S}}}{\braket{B_i^2}-\braket{B_i}^2}+\text{O} \lc \sqrt{G_N} \rc\\
    &= B_i \ket{0,\mathcal{S}} + \text{O} \lc \sqrt{G_N} \rc.
    \end{split}
\end{equation}
Thus we can clearly see that $\braket{A_j C_i} = \braket{A_j B_i} + \text{O} \lc \sqrt{G_N} \rc$ from here.

\subsection*{Boundedness of $C_i$}
Let us define the orthonormal states $\ket{B_i^\perp,\{s\}} \equiv \frac{1}{\beta_i}\enc{1-\pvacex}B_i\ket{0,\{s\}}$, where $\beta_i\equiv \sqrt{\braket{B_i^2} - \braket{B_i}^2}$. Then $C_i$ can we written as
\begin{equation}
\begin{split}
    C_i = \int \lc \prod_{l,m} ds_{l,m} \rc\,  & \Big(\ket{0,\{s\}}\bra{0,\{s\}} + \beta_i \enc{\ket{0,\{s\}}\bra{B_i^\perp,\{s\}}+\ket{B_i^\perp,\{s\}}\bra{0,\{s\}}}  \\
     &- \braket{B_i} \ket{B_i^\perp,\{s\}}\bra{B_i^\perp,\{s\}} \Big) .
\end{split}
\end{equation}
In terms of the orthonormal basis $\{\ket{0,\{s\}, \ket{B_i^\perp}\,\{s\}}\}$, it takes the form 
\begin{equation}
    C_i = \begin{pmatrix}
    \braket{B_i} & \beta_i \\
    \beta_i & -\braket{B_i}
    \end{pmatrix} \otimes \mathbbm{1}  + \text{O} \lc \sqrt{G_N} \rc
\end{equation}
where the $\otimes \mathbbm{1}$ stands for $C_i$'s identity action on supertranslation sectors. The eigenvalues of $C_i$ are $\pm\sqrt{\braket{B_i^2}} + \text{O} \lc \sqrt{G_N} \rc$ and hence the norm is
\begin{equation}
    \norm{C_i}^2 = \braket{B_i^2} + \text{O} \lc \sqrt{G_N} \rc <1.
\end{equation}

\section{Proof of existence of $C_i$ of \S \ref{422}}
\label{sec:general_rs_proof}
Let's split the boundary observable as $C=\Cex+\Cdelta$ where $\Cex$ contains $\pvacex$ and $\Cdelta$ contains $\delta \mathcal{P}$ (CHSH label is suppressed). We know that the $\Cex$ part gives the desired correlators and is bounded in a desired way by constructing a boundary $Q_i$ such that $Q_i\vacket=B_i\vacket$, which Reeh-Schlieder guarantees can always be done. We need to show that $\braket{A_j\Cdelta_i}$ can be made arbitrarily small. We have
\begin{equation}
    \beta_i^2 \braket{A_j\Cdelta_i} = \braket{B_i^2} \underbrace{\bra{\Omega}A_j\delta\mathcal{P}}_{\text{energy$<\delta$}}Q_i^\dagger\vacket - \braket{B_i} \underbrace{\bra{\Omega}A_jQ_i\delta\mathcal{P}}_{\text{energy$<\delta$}}Q_i^\dagger\vacket.
\end{equation}
Both of these terms can be interpreted as the inner product of $Q_i^\dagger\vacket$ and a bra which contains excitations on $\bra{\Omega}$ with energy less than $\delta$. The latter is a linear combination of bras of the kind $\bra{\Omega}a_{\omega_1}a_{\omega_2}...a_{\omega_n}$ such that $\sum_{j}\omega_j<\delta$. Here we have suppressed the $l$ label of the global annihilators because they don't contribute to energy. Each of these terms is
\begin{align}
    \begin{split}
        \braket{\Omega|a_{\omega_1}a_{\omega_2}...a_{\omega_n}Q_i^\dagger|\Omega}=
        \braket{\Omega|a_{\omega_1}...a_{\omega_{j-1}}a_{\omega_{j+1}}...a_{\omega_n}[a_{\omega_j},Q^\dagger_i]|\Omega},
    \end{split}
\end{align}
where $\omega_j$ is any of the $n$ different energies. So these terms can be made arbitrarily small individually if we can guarantee
\begin{align}
    [Q_i, a^\dagger_{\omega}] \approx 0 \qquad \forall \; 0<\omega<\delta. 
    \label{eq:ir_deletion_condition}
\end{align}
where $\approx$ has been used to mean ``arbitrarily close to''. This condition requires that $Q_i$ in addition to satisfying
\begin{equation}
    Q_i\vacket\approx B_i\vacket,
    \label{eq:q_condition}
\end{equation}
needs to be constructed in a way such that it contains (arbitrarily) small contribution from $a_{\omega,l}$ for $\omega<\delta$. To make this condition more precise we inspect how smearing of the field operator translates into smearing of creation and annihilation operators in energy domain. Consider a smearing of the kind $\phi_f=\int dt f(t)\phi(t)$. We have suppressed the position argument of both the field $\phi$ and the smearing function $f$ for simplicity. This decomposes as,
\begin{align}
    \begin{split}
        \phi_f &= \int dt\, f(t) \phi(t) \\
        &=\int dt\, f(t) \int_0^\infty d\omega \, \enc{e^{-i\omega t}a_\omega + e^{i\omega t}a^\dagger_\omega} \\
        &=\int_{0}^\infty d\omega\, \enc{\hat{f}(\omega)a^\dagger_{\omega}+ \hat{f}(-\omega)a_{\omega}} \\
        &= a^\dagger_{\hat{f}^+} + a_{\hat{f}^-},
    \end{split}
\end{align}
where the hats represent time domain Fourier transforms, the subscripts on mode operators denote the frequency space smearing: $a^\dagger_{\hat{f}^\pm}\equiv \int_{0}^\infty d\omega\,\hat{f}^\pm(\omega)a^\dagger_{\omega}$ and $\hat{f}^{\pm}(\omega)=\hat{f}(\pm\omega)$. Again, we have suppressed the sum over spherical mode information $l$ for the creation and annihilation operators for brevity. Evidently, creation and annihilation operators are weighted by the positive and negative Fourier modes of $f(t)$ respectively.

Now we make the condition set on $Q_i$ more precise. Let $\mathcal{A}$ be the algebra generated by all $\phi$ smearings in region $C$. Also, let $\mathcal{A}_{\delta,\theta}$ be the subset of $\mathcal{A}$ containing operators of the kind $\phi_{f_1} + \phi_{f_2}\phi_{f_3}+...$  such that $\sum_i\int_{0}^\delta d\omega\, \abs{\hat{f}_i^{-}(\omega)}^2<\theta$. In simple terms, $\mathcal{A}_{\delta,\theta}$ is a subset of $\mathcal{A}$ in which all elements obey \cref{eq:ir_deletion_condition} up to precision $\theta$ (this may not be the maximal subset with this property). We shall argue for the existence of a $Q_i$ obeying both \cref{eq:ir_deletion_condition,eq:q_condition} by showing that $\mathcal{A}_{\delta,\theta}\vacket$ is dense in the entire Hilbert space $\mathcal{H}$ which is generated by field operations on $\vacket$ for any $\theta>0$ however small. Notice that the denseness of $\mathcal{A}\vacket$ in $\mathcal{H}$ is just the statement of the Reeh-Schlieder theorem and hence the denseness of $\mathcal{A}_{\delta,\theta}\vacket$ in $\mathcal{H}$ is not too surprising.

To simplify things a little, we take $\mathcal{A}$ to be the algebra of all operator smearings with support in some time band $[0,\epsilon]$ (as a simplified model of region $C$). We also simplify the definition of $\mathcal{A}_{\delta,\theta}$ accordingly. Now, let us split the full Hilbert space into particle number sectors as $\mathcal{H}=\bigoplus_{n=0}^\infty\mathcal{H}_n$, where the $n$-particle sector $\mathcal{H}_n$ contains states like $a^\dagger_{\hat{f}_1}...a^\dagger_{\hat{f}_n}\vacket$. We shall show that $\mathcal{A}_{\delta,\theta}\vacket$ is dense in both even and odd particle sectors by induction.

\textbf{Even sector:}
Consider the hypothesis: $\mathcal{A}_{\delta,\theta}$ is dense in $\mathcal{H}_n$. By axiom, the global identity operator $\mathbb{I}$ exists in $\mathcal{A}_{\delta,\theta}$ ($\mathcal{A}$ is a von Neumann algebra), and hence $\mathcal{A}_{\delta,\theta}\vacket$ contains $\vacket$. Therefore $\mathcal{A}_{\delta,\theta}\vacket$ is dense in $\mathcal{H}_0$, i.e. the hypothesis is true for $n=0$. Consider a general $(2n+2)$ particle term $a^\dagger_{\hat{f}_1...}a^\dagger_{\hat{f}_{2n+2}}\vacket$. Because we have all the $\hat{f}_i(\omega)$ at our disposal, by the corollary stated and proved in \cref{sec:fourier_proof}, we can construct a $g_{i}(t)$ with support in $[0,\epsilon]$ such that $\hat{g}^+_i(\omega) \approx \hat{f}_i(\omega)$ and $\hat{g}^-_i(\omega)\approx0$ for $0<\omega<\delta$. This gives
\begin{equation}
    \phi_{g_1} \phi_{g_2} ... \phi_{g_{2n+2}} \vacket \approx a^\dagger_{\hat{f}_1} a^\dagger_{\hat{f}_2} ... a^\dagger_{\hat{f}_{2n+2}} \vacket + \enc{\mathcal{H}_{2n} \text{ term}} +...+\enc{\mathcal{H}_{2} \text{ term}} + \enc{\mathcal{H}_0\text{ term}}.
\end{equation}
The first term on the RHS is the one we need to approximate, but other lower particle number terms show up due to the non-commutativity of creation and annihilation operators. If the hypothesis is true for $n=2,4,...,2n$, then these residual terms are also limit points of $\mathcal{A}_{\delta,\theta}\vacket$ and can be cancelled off to any precision by summoning a state from $\mathcal{A}_{\delta,\theta}\vacket$. But this means the (2n+2) particle term is also a limit point of $\mathcal{A}_{\delta,\theta}\vacket$. Hence the hypothesis is true for all even $n$ (including $0$) and we have proved by strong induction that $\mathcal{A}_{\delta,\theta}\vacket$ is dense in all even number particle sectors.

\textbf{Odd sector}:
Consider a state $a^\dagger_{\hat{f}}\vacket \in \mathcal{H}_1$. Just like in the even case, $\hat{f}_1(\omega)$ lets us construct $g_1(t)$ with support in $[0,\epsilon]$ such that $\hat{g}^+_1(\omega) \approx \hat{f}_1(\omega)$ and $\hat{g}^-_1(\omega)\approx0$ for $0<\omega<\delta$. Then we have 
\begin{equation}
    \phi_g \vacket \approx a^\dagger_{\hat{f}_1} \vacket.
\end{equation}
Since $\hat{f}_1(\omega)$ was a general smearing function, we know $\mathcal{A}_{\delta,\theta} \vacket$ is dense in $\mathcal{H}_1$ and the hypothesis is true for $n=1$. Tracing the exact same inductive steps as the above case, we obtain that $\mathcal{A}_{\delta,\theta} \vacket$ is dense for all odd $n$.

This concludes the proof for denseness of $\mathcal{A}_{\delta,\theta}\vacket$ in $\mathcal{H}$, and hence also the proof for the existence of a $Q_i$ localised in region $C$ and satisfying both \cref{eq:ir_deletion_condition,eq:q_condition}.
\subsection{Positive Fourier mode reconstruction}
\label{sec:fourier_proof}
\textbf{Lemma:} Given $\delta,\epsilon>0$, the space of $L^2\mathbb{R}$ functions with support in $[0,\epsilon]$ is dense in $L^2\mathbb{R}$ under the norm defined by $\norm{f}^2_\delta = \int_{-\delta}^\infty d\omega\, \abs{\hat{f}(\omega)}^2$, where $\hat{f}(\omega)=\int_{\mathbb{R}}\tfrac{dt}{2\pi}\, f(t)e^{i\omega t}$. More explicitly, given a function $f\in L^2\mathbb{R}$, and $\delta,\epsilon,w>0$, there exists a function $g\in L^2[0,\epsilon]$ such that
\begin{equation}
    r\equiv\int_{-\delta}^{\infty} d\omega\, \abs{\hat{f}(\omega)-\hat{g}(\omega)}^2 < w.
\end{equation}
\textbf{Proof:}
Let $P_\epsilon$ be the projector onto the space of all functions supported in $[0,\epsilon]$, and $P_{-\delta}$ be the projector onto the space of all functions which contain no Fourier modes in the range $(-\infty,-\delta)$. The quantity in question, $r=\int_{-\delta}^{\infty} d\omega\, \abs{\hat{f}(\omega)-\hat{g}(\omega)}^2$ is manifestly equal to $\tfrac{1}{2\pi}\norm{P_{-\delta}f-P_{-\delta}g}^2$, where $\norm{}$ is the standard $L^2$ norm. Since, we need to show the existence of a $g\in P_\epsilon L^2\mathbb{R}$ such that $r$ can be made arbitrarily small, it is enough to show that the subspace $P_{-\delta}P_{\epsilon}L^2\mathbb{R}$ is dense in $P_{-\delta}L^2\mathbb{R}$. Let $\Cc$ be the subspace of all smooth functions with compact support in $L^2\mathbb{R}$. This subspace happens to be dense in $\LtwoR$. Let us first show the denseness of $P_{-\delta}P_\epsilon \LtwoR$ in the subspace $P_{-\delta}\Cc$. We shall show this by contradiction.

Let $P_{-\delta}P_{\epsilon}L^2\mathbb{R}$ \textit{not} be dense in $P_{-\delta}\Cc$. Then there exists a non-zero function $\chi\in P_{-\delta}\Cc$ such that $(\phi,\chi)=\int_{\mathbb{R}}dt\, \phi^*(t) \chi (t)=0$ for all $\phi \in P_{-\delta}P_{\epsilon}\LtwoR$. So,
\begin{align}
    \begin{split}
        (P_{-\delta}P_\epsilon\psi, \chi) &= 0 \quad \forall \; \psi \in L^2\mathbb{R} \\
        \Rightarrow (\psi,P_\epsilon P_{-\delta} \chi) &= 0 \quad \because \; \text{both projectors are hermitian} \\
        \Rightarrow(\psi,P_\epsilon \chi) &= 0 \quad \because \; P_{-\delta}\chi = \chi \\
        \Rightarrow P_\epsilon \chi &= 0 \quad \because \; \text{$\psi$ is arbitrary} \\
        \Rightarrow \chi(t) &= 0 \quad \forall \: t \in [0,\epsilon].
    \end{split}
\end{align}
Now, since $\chi(t)$ is a smooth function on $\mathbb{R}$, identically vanishing over the interval $[0,\epsilon]$ means while we Taylor expand it around some point in this interval, say $0$, all the Taylor coefficients turn out to be $0$. $\chi(t)$ therefore vanishes identically all throughout the real line. This is in contradiction to the hypothesis that $\chi(t)$ is non-zero. Hence we have shown that $P_{-\delta}P_\epsilon\LtwoR$ is dense in $P_{-\delta}\Cc$. On the other hand, $P_{-\delta}\Cc$ is dense in $P_{-\delta}\LtwoR$ because $\Cc$ is dense in $L^2\mathbb{R}$. Hence, by transitivity of denseness of topological spaces, we have shown $P_{-\delta}P_\epsilon\LtwoR$ is dense in $P_{-\delta}\LtwoR$. 
\\
\textbf{Corollary:} Given a function $f\in L^2\mathbb{R}$, and $\delta,\epsilon,w>0$, there exists a function $g\in L^2[0,\epsilon]$ such that
\begin{equation}
    \int_{-\delta}^{0} d\omega\, \abs{\hat{g}(\omega)}^2+\int_{0}^{\infty} d\omega\, \abs{\hat{f}(\omega)-\hat{g}(\omega)}^2 < w.
\end{equation}
In other words, for any function $f(t)$, there exists a $g(t)$ supported in $[0,\epsilon]$ such that it approximates $f(t)$ in the positive Fourier modes with its modes in the $[-\delta,0)$ range suppressed to arbitrary precision.
\\ 
\textbf{Proof:} Given $f\in L^2\mathbb{R}$, construct $f_1\in L^2\mathbb{R}$ by deleting its Fourier modes in the $[-\delta,0)$ range. That is 
\begin{equation}
    \hat{f}_1(\omega) = \begin{cases} 0 \quad &\text{if} \quad \omega \in [-\delta,0) \\ 
    \hat{f}(\omega) \quad &\text{if} \quad \omega \notin [-\delta,0) 
    \end{cases}.
\end{equation}
Now applying the above lemma to $f_1(t)$ instead of $f(t)$ proves the existence of the desired $g(t)$.
\bibliographystyle{JHEP}
\bibliography{citation.bib}
\end{document}